\documentclass[fleqn,usenatbib]{mnras}

\usepackage{newtxtext,newtxmath}

\usepackage[T1]{fontenc}
\usepackage{ae,aecompl}

\usepackage{graphicx}	
\usepackage{epstopdf}
\usepackage{amsmath}	
\usepackage{amssymb}	
\usepackage{float}
\usepackage{multirow}

\title[Deblending and Classifying with Mask R-CNN]{Deblending and Classifying Astronomical Sources with Mask R-CNN Deep Learning}

\author[C. J. Burke et al.]{
Colin J. Burke,$^{1,2}$\thanks{E-mail: colinjb2@illinois.edu}
Patrick D. Aleo,$^{1,3}$\thanks{E-mail: paleo2@illinois.edu}
Yu-Ching Chen,$^{1,2}$
Xin Liu,$^{1,2}$
\newauthor
~John R. Peterson,$^{4}$
Glenn H. Sembroski,$^{4}$
Joshua Yao-Yu Lin$^{5}$
\\
$^{1}$Department of Astronomy, University of Illinois at Urbana-Champaign, 1002 West Green Street, Urbana, IL 61801, USA\\
$^{2}$National Center for Supercomputing Applications, 1205 West Clark Street, Urbana, IL 61801, USA\\
$^{3}$Advanced Visualization Laboratory, National Center for Supercomputing Applications, 1205 West Clark Street, Urbana, IL 61801, USA\\
$^{4}$Department of Physics and Astronomy, Purdue University 525 Northwestern Avenue, West Lafayette, IN 47907, USA\\
$^{5}$Department of Physics, University of Illinois at Urbana-Champaign, 1110 West Green Street, Urbana, IL 61801, USA
}

\date{Accepted XXX. Received YYY; in original form ZZZ}

\pubyear{2019}

\begin{document}
\label{firstpage}
\pagerange{\pageref{firstpage}--\pageref{lastpage}}
\maketitle

\begin{abstract}
We apply a new deep learning technique to detect, classify, and deblend sources in multi-band astronomical images. We train and evaluate the performance of an artificial neural network built on the Mask R-CNN image processing framework, a general code for efficient object detection, classification, and instance segmentation. After evaluating the performance of our network against simulated ground truth images for star and galaxy classes, we find a precision of 92\% at 80\% recall for stars and a precision of 98\% at 80\% recall for galaxies in a typical field with $\sim30$ galaxies/arcmin$^2$. We investigate the deblending capability of our code, and find that clean deblends are handled robustly during object masking, even for significantly blended sources. This technique, or extensions using similar network architectures, may be applied to current and future deep imaging surveys such as LSST and WFIRST. Our code, \texttt{Astro R-CNN}, is publicly available at \url{https://github.com/burke86/astro_rcnn}.
\end{abstract}




\section{Introduction}
\label{sec:intro}

The next generation of astronomical surveys such as the Large Synoptic Survey Telescope \citep[LSST;][]{Ivezic2019}, the Wide-Field Infrared Survey Telescope  \citep[WFIRST;][]{Spergel13}, and Euclid \citep{Amiaux12} will produce unprecedented amounts of imaging data throughout the 2020s. This quickly-approaching era demands efficient, uniform, and robust techniques to detect, classify, and analyze sources in images.

The task of star/galaxy classification is a long-standing problem in astronomy, dating back to the likes of \cite{Messier1781}. Until recently, a technique known as morphological separation \citep{Sebok1979, Valdes1982} was the popular choice, which involved a simple assumption: galaxies are resolved sources, and stars point sources. \cite{Sebok1979} and \cite{Valdes1982} pioneered a Bayesian approach focusing on classifying objects by maximizing the probability of object class models matching the observed pixel intensity distributions. In contrast, \cite{Jarvis81} use a parametric method, where clustering of data points of measured pixel intensity distribution determines the classification.

Next-generation ground-based surveys, such as LSST, will detect numerous unresolved and marginally-resolved galaxies, particularly near the photometric limit. In this regime, a strictly morphology-based approach will not be able to consistently differentiate between stars and galaxies \citep{Kim2015}. Thus, several studies have introduced machine learning methods such as decision trees \citep{Vasconellos2011}, a blend of different learning approaches \citep{Kim2015,Soumagnac2015}, and deep learning \citep{Serra-Ricart1996, Kim2017}. See \cite{Cheng2019} for an overview and comparison of different machine learning galaxy classification techniques.

As the both the sensitivity and depth of surveys increase, we will encounter larger numbers of blended (overlapping) sources due to line-of-sight projection or source interaction (i.e. galaxy-galaxy mergers). The detection of both fainter galaxies and more extended regions of objects will increase the probability of blending. If blends are not identified, they will bias results from pipelines that assume object isolation. Some important examples include photometric redshifts \citep{Boucaud19} and weak lensing \citep{Arneson2013}. Once LSST begins its survey, efficient deblending techniques will be a necessity, and thus been recognized a high priority in the field.

Current estimates place the fraction of significantly blended galaxies (3" center-to-center distance) in LSST images at roughly $50\%$ \citep{Dawson16,Dawson14,Chang13}. \cite{Chang13} estimates roughly $10\%$ of galaxies to be blends with a 1" center-to-center distance in a typical region of the sky (around 37 galaxies per arcmin$^2$). In all, if effective deblending algorithms are not put in place during the ten-year LSST survey, roughly 200 million galaxies could be discarded each year according to \cite{Reiman19}. \par

Even in current surveys, such as the Dark Energy Survey \citep[DES;][]{Abbott2018}, crowded fields are challenging for the current pipeline. For this reason, a deblending code was developed by \citet{Zhang2015} for DES images of clusters of galaxies. See \cite{Sevilla-Noarbe2018} for a summary of star/galaxy classification techniques used in DES. The Hyper Suprime-Cam (HSC) Subaru Strategic Program \citep{Aihara2018} also suffers from poor photometry in crowded fields due to significant blending \citep{Huang2018,Aihara2019}. Several existing codes for source detection and classification incorporate simple deblending algorithms into their frameworks, such as $\texttt{FOCAS}$ \citep{Jarvis81}, $\texttt{SExtractor}$ \citep{Bertin96}, and $\texttt{NEXT}$ \citep{Andreon00}. However, these approaches are highly sensitive to configuration parameters, such as the density of sources in the field. Additionally, these codes can be inefficient compared to machine learning approaches.

Recently, new algorithmic techniques have been developed in the context of LSST such as the work of \citet{Lupton2014} and $\texttt{SCARLET}$ \citep{Melchior2018}. Similar to \texttt{Astro R-CNN}, $\texttt{SCARLET}$ takes advantage of multi-band imaging when there are overlapping sources. It achieves this by utilizing non-negative matrix factorization, where one matrix factor stores the source color and the other spatial shape information. Then, the likelihood function is minimized with respect to the inherent non-negativity constraints via a proximal gradient descent update.

The recent work of \cite{Hausen2019} combined deep learning semantic segmentation with mask separation algorithms to classify and deblend galaxies in the Hubble Space Telescope images. Their code, \texttt{Morpheus}, is based on the U-Net \citep{Ronneberger2015} architecture, and like \texttt{Astro R-CNN}, performs pixel-level classifications. Further, both codes uniquely identify source and background pixels, allowing for a singular, cohesive analysis of object detection and classification. Although, \texttt{Morpheus} utilizes user-supplied segmentation maps and utilizes the classic watershed transform algorithm \citep{Couprie1997} to separate source and background pixels. These source pixels are subsequently classified in terms of their morphology (background, disk, spheroid, irregular, point source/compact).

Some recent works have used more experimental artificial neutral network techniques for deblending and shown promising results \citep{Reiman19,Boucaud19,Gonzalez2018}. \cite{Zhang2019} used deep learning to mask cosmic rays. With recent ubiquity of powerful machines with multiple graphics processing units (GPUs) and a plethora of machine learning libraries, these techniques have never been more accessible or appealing.

Many classification and deblending codes operate monochromatically, not taking into account source color gradient information. In the context of DES and LSST, their uniform, multi-band data should be exploited to assist with source classification and deblending. In addition, the source spectral energy distribution (SED) information can be used to help discriminate between stars and unresolved galaxies or quasars. Multi-band images can also be used with convolutional neural networks to estimate photometric redshifts of galaxies and quasars \citep{Pasquet2019,DIsanto2018}.

In this work, we develop a new deep learning method based on the Mask Region-based Convolutional Neural Network (Mask R-CNN) framework \citep{He17} to perform all tasks of source detection classification, and deblending in a single machine learning framework. We train and validate our network using simulated images and catalogs, and test its performance using DECam Legacy Survey \citep[DECaLS;][]{Dey2019} images of a crowded field. After training, this method is extremely efficient, detecting, classfying, and segmenting a 512 $\times$ 512 pixel$^2$ image in $\lesssim$100 milliseconds using a single NVIDIA Tesla V100 GPU. This work may be extended to other telescopes and surveys using transfer learning. The code is open source and available at the $\texttt{Astro R-CNN}$ GitHub repository \citep{repo}.

This paper is organized as follows. In \S\ref{sec:network}, we introduce the Mask R-CNN framework and describe the architecture of our implementation. We also explain our training procedure using transfer learning and simulated images. In \S\ref{sec:results}, we present the results of our trained neural network, validate the results using the simulated catalog as a ground truth, and evaluate its performance. We present results using real DECam images. In \S\ref{sec:discussion}, we discuss the implications of our method and its benefits and drawbacks compared to existing work. In  \S\ref{sec:conclusions}, we summarize our findings and conclude.

\section{Network \& Training}
\label{sec:network}

Object detection, classification, and instance segmentation is an active area of research in the field of computer vision. There are several machine learning -based solutions that perform semantic/instance segmentation, such as YOLO \citep{Redmon2015}, YOLACT \citep{Bolya2019}, PANet \citep{Liu2018}, and TernausNet \citep{Iglovikov2018}.

Recently, \cite{He17} developed a novel and general framework for instance segmentation called Mask R-CNN. This extends earlier work of Fast/Faster R-CNN \citep{Girshick2015,Ren2015}, a deep convolutional neural network for classification and bounding box recognition. Mask R-CNN adds a parallel branch for instance segmentation (Fig.~\ref{fig:maskrcnn_arch}). The Mask R-CNN framework is highly efficient and robust to occlusion. Several recent works have applied Mask R-CNN to different fields from cellular biology \citep{Tsai2019} to remote sensing \citep{Zhang2018}. We select this framework for its recent ubiquity and maturity. Even more recent works have built-upon Mask R-CNN \citep[e.g. Mask Scoring R-CNN;][]{Huang2019} and \cite{Zimmermann2018}, but are not explored in this work.

In this section, we describe our Mask R-CNN implementation for the purpose of star/galaxy detection, classification, and deblending. We also outline our training, validation, and test dataset generation using simulated images. Our training procedure is a form of supervised learning where simulated images and labeled masks are used to train the network. We describe this process in detail below.

\subsection{Implementation}

The code developed in this work extends the Python language implementation of Mask R-CNN from \cite{Matterport17}. This code is built on the Keras library \citep{Chollet2015} using a TensorFlow \citep{Abadi2016} backend. We allow multi-band flexible image transport system (FITS) files \citep{Pence2010} as image input during both the training and detection (inference) modes. The final segmentation masks are saved as multi-extension FITS files. The problem of source detection, classification, and deblending in astronomical surveys is well-suited to the Mask R-CNN framework which performs all tasks in one cohesive package.

\begin{figure}
	\includegraphics[width=\columnwidth]{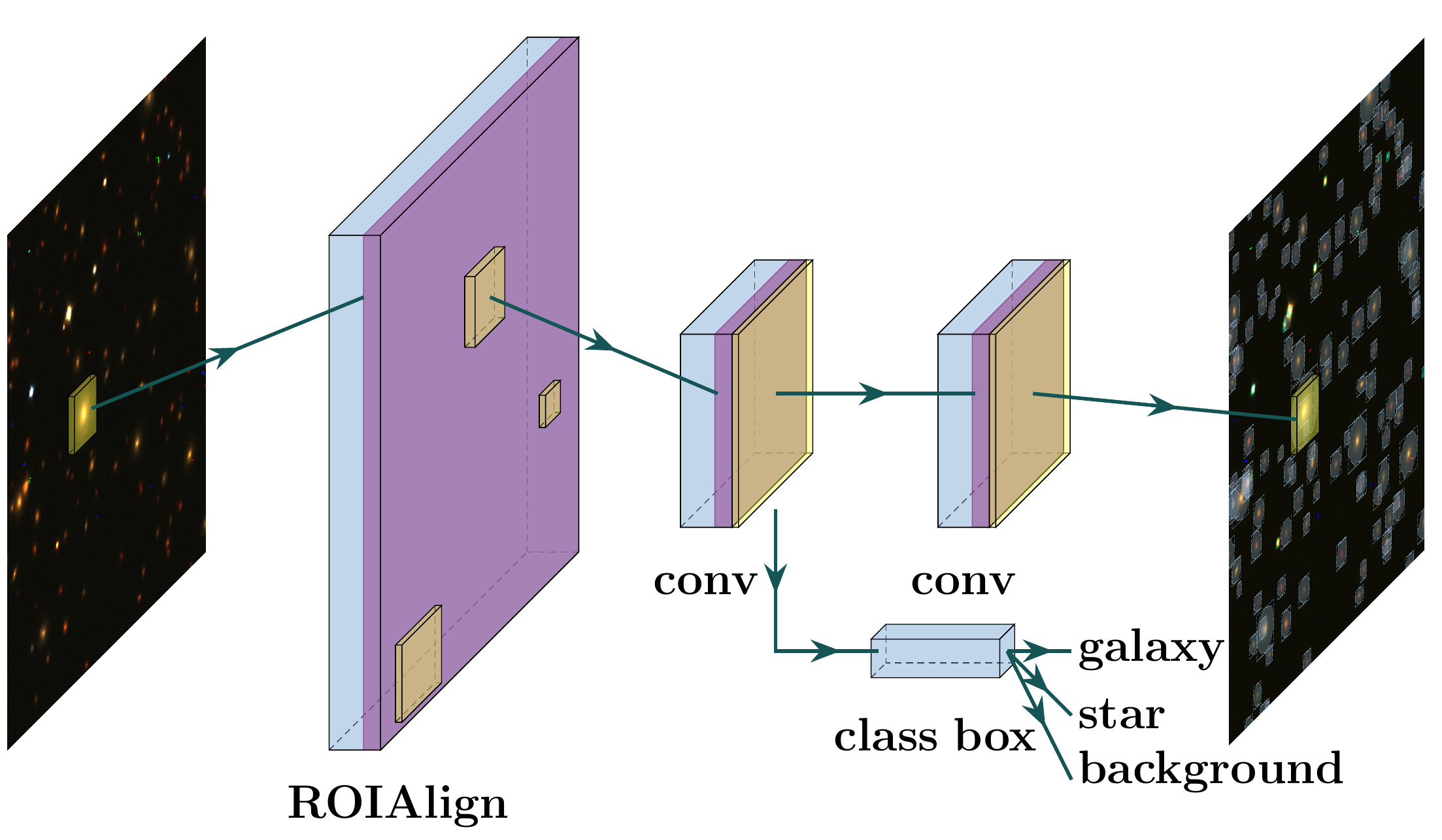}
    \caption{A simple schematic of our Mask R-CNN framework for instance segmentation, based on Figure 1 of \protect\cite{He17}. The feature map in RoIAlign has a few region of interest (RoI) examples, shown in yellow, where one galaxy is propagated through the network. In the convolutional layers this RoI is a fixed size feature map. Mask R-CNN extends Faster R-CNN by parallelizing the branch for segmentation mask prediction with the branch for classification and bounding box regression.}
    \label{fig:maskrcnn_arch}
\end{figure}

Our Mask R-CNN implementation uses the standard 101-layer deep residual network \citep[ResNet-101;][]{He2016} architecture as its backbone. On its own, ResNet-101 is a feature extractor, wherein earlier layers detect low-level features (e.g. corners or edges) and later levels detect high-level features (e.g. galaxies or stars) using residual learning. By using shortcut connections in 3-layer deep residual blocks, ResNet-101 is able to solve the \emph{degrading accuracy problem}: with increasing network depth, accuracy becomes saturated and degrades quickly \citep{Bengio1994, He2016}. This occurs when back-propagating the gradient, where repeated multiplications involving small weights tend to create smaller and smaller gradients to the point of ineffectiveness \citep{Huang2016}. This is referred to as the \emph{vanishing gradient problem}.

As with Faster R-CNN, Mask R-CNN is a feature pyramid network (FPN) which defines hierarchically-sized \emph{anchors} for multi-scale object recognition. This FPN adds a second pyramid network that allows high level features to be passed down to lower levels and vice-versa, so that any layer has an awareness of both low- and high-end features \citep{Lin17}. With a feature map at each level of the second pyramid (instead of a single backbone feature in the top layer of the first pyramid), the one most appropriate for the size of the object is chosen at runtime, ultimately enabling better feature extraction \citep{Matterport17, Lin17}.

\begin{figure}
	\includegraphics[width=\columnwidth]{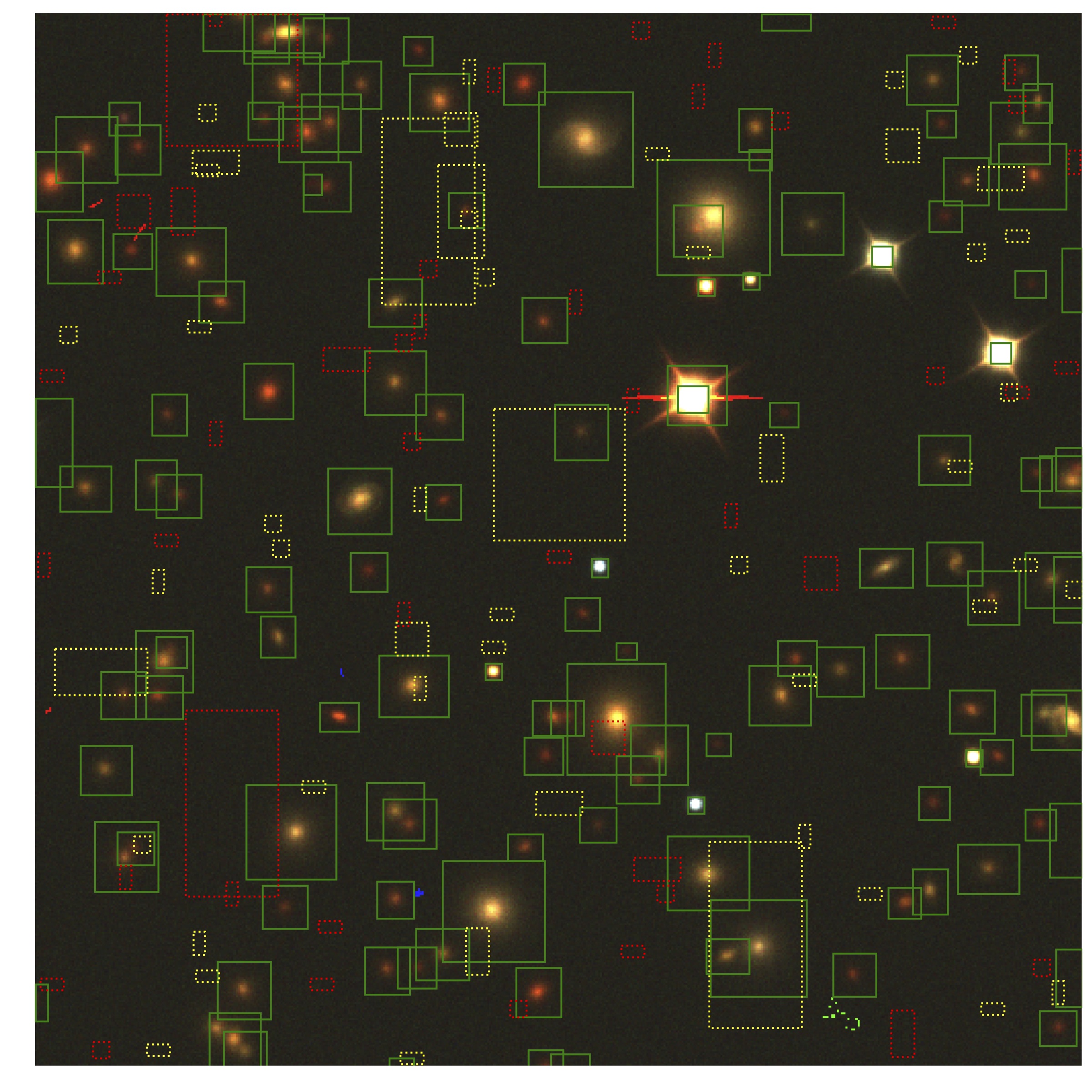}
    \caption{Example of a simulated DECam image cutout with RoI bounding boxes overlayed. Positive RoIs (solid green) correspond to a region where a significant detection is identified. For clarity, only 50 neutral (dashed yellow) and 50 negative (dashed red) RoIs are shown, and the RoI bounding boxes are enlarged. Note that the RPN scans over the backbone feature map and not the image directly as shown here; this is done for illustrative purposes to develop an intuition of how the RPN will identify features for extraction.}
    \label{fig:roi}
\end{figure}

Once the appropriate backbone feature map is chosen, this is fed to the region proposal network (RPN) for scanning \citep{Ren2015}. The RPN is a neural network which scans in a sliding window fashion over thousands of anchors---overlapping areas of different sizes and aspect ratios---to generate two outputs for each anchor: an associated anchor class and bounding box refinement. For anchor class, the anchor will be assigned as
\textit{foreground/positive} or \textit{background/negative}, where the former suggests the existence of an object contained within it, and the latter does not (a third type, \textit{neutral}, does not contribute to training). If assigned positive, the bounding box refinement will generate a suggested shifted bounding box to place the object at its center. These predictions most likely to contain objects are called proposals and sent to the next stage as regions of interest (RoIs; Fig.~\ref{fig:roi}).

\begin{figure*}
	\includegraphics[width=\textwidth]{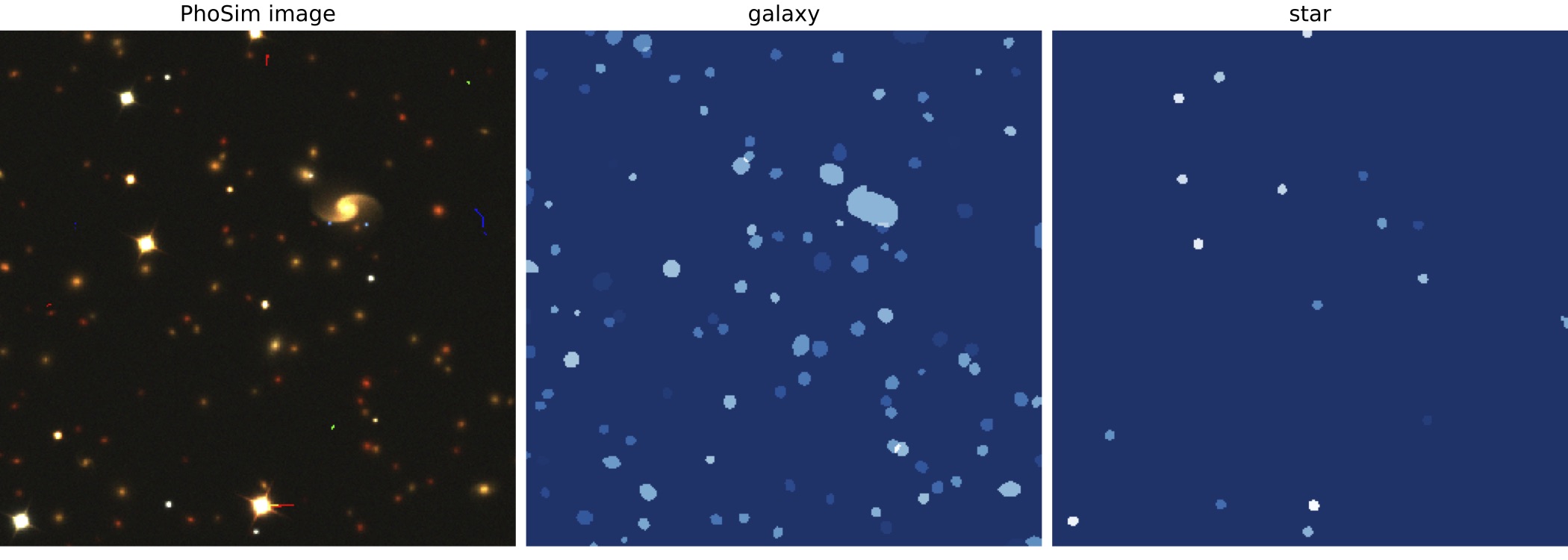}
    \caption{Typical example of PhoSim DECam training data. The simulated color image of a crowded field (\emph{left panel}) and masks corresponding to both galaxy (\emph{center panel}) and star (\emph{right panel}) classes are pictured. Note the realism of the simulated images: the distribution and colors of sources, complex galaxy morphology, diffraction and blooming (most noticeably for the bright stars), bleed trails, and cosmic rays are clearly visible.}
    \label{fig:example}
\end{figure*}

In this work we implement three RoIs classifications: star, galaxy, or background. However, the background class is trivial and will be ignored in our analysis. This essentially makes our model a simple binary classifier. Now, each RoI is assigned another bounding box. Unfortunately, RoI boxes are likely to be of all different sizes, which confuses classifiers. To circumvent this, RoIAlign samples the feature maps and applies bilinear interpolation\footnote{The code we adopted uses TensorFlow's \emph{crop\_and\_resize} function as a numerically-efficient approximation.} such that there is a fixed input size (represented by the first layer after RoIAlign in Fig~\ref{fig:maskrcnn_arch}). RoIs are aligned with the RoIAlign layer and piped through the CNN, a process founded by \cite{He17} to improve AP scores over the standard RoIPool \citep{Girshick2015} due to its preservation of exact spatial locations critical to feature extraction \citep{He17}.

In final instance segmentation (masking) stage, each pixel is assigned a class which can be used to mask sources and obtain (deblended) cutouts from full-scale images. Mask shapes are stored at bounding box positions using the ``mini$~$masks'' feature to optimize the training. We use a fixed image size of 512$\times$512 (image sub-region) $\times$3 (\textit{g},\textit{r},\textit{z} bands) pixel$^3$ to simplify the training. These image sub-regions are used in our extension of Mask R-CNN, and the output can be tiled to full-scale images or mosaics.

\subsection{Training Set}
\label{sec:trainingset}

A common problem of artificial neural network techniques is the lack of a sufficiently-sized, uniform, and unbiased training set. For example, one may wish to train on astronomical sources which are too faint for current surveys or too rarely observed. We alleviate these problems by using simulated images of crowded extragalactic fields. Importantly, using simulated images gives us a large training set of a known catalog of stars and galaxies. This data is used as a truth to test our network's performance without the misclassification errors in real catalogs.

We invoke The Photon Simulator \citep[PhoSim;][]{Peterson2015} to simulate DECam images. PhoSim is an \textit{ab initio} photon Monte Carlo code originally developed for LSST. We use PhoSim with the Blanco 4-m DECam telescope/instrument options adapted from \citet{Flaugher2015}. The PhoSim DECam implementation includes a full description of the optical prescription and focal plane, as described by \cite{Cheng2017}. PhoSim can quickly simulate images of pseudorandom star and galaxy catalogs under a distribution of typical observing conditions. PhoSim includes all relevant physics of the atmosphere, telescope and camera optics, and detector. Below, we detail our procedure for generating a large, uniform, and realistic training set of PhoSim DECam images.

Simulated galaxies are described using PhoSim's \verb$sersicComplex$ three-dimensional galaxy model. This model includes three-dimensional ellipsodial S\'ersic profiles \citep{Sersic1963} for both the bulge and disk morphology, along with additional parameters for describing irregular knots and spiral structure and their three-dimensional orientations. By sampling these parameters from realistic distributions, PhoSim can simulate images with spiral, elliptical, or irregular galaxies. The approximate number density and population of galaxies is derived from the cosmic star-formation history described in \cite{Madau2014}. Galaxies are given random, typical SEDs for the bulge and disk, derived from \citet{Molla2009}. A simple model accounting for the different stellar ages and metallicities of the bulge and disk is considered \citep{Peletier1996,Gallazzi2005}.

Milky Way stars are simulated as point sources whose number density distribution varies as a function of galactic latitude. The stellar population follows the initial mass function described in \cite{Kroupa2001}. Stellar SEDs are derived from \citet{Castelli2003} and \citet{Kurucz1993}. Stellar metallicities and atmosphere parameters are derived from \citet{Prieto2004} and \citet{Prugniel2011}.

Next, we outline our procedure for generating the training set data. The only additional change made to PhoSim is to use an increased number density of galaxies to simulate a more dense extragalactic field, such as a cluster of galaxies. We use a 4$\times$ overdensity of galaxies which works out to about $30$ galaxies/arcmin$^2$ in our images. We execute PhoSim in two stages, which are outlined as follows:
\begin{enumerate}
  \item Simulate 512$\times$512 pixel$^2$ ($\approx$ 5 arcmin$^2$) DECam images in 3 bands (\textit{g},\textit{r},\textit{z}) of a pseudorandom crowded extragalactic field with a 150 s integration time. We simulate stars and galaxies according to PhoSim's realistic distributions between the typical magnitude range $12<g<23$. This integration time and limiting magnitude roughly approximates typical DECaLs data release 7 coadds \citep{Dey2019}. Given the various observing conditions, the typical minimum signal-to-noise ratio limit for sources in our images is $S/N\sim2$ Fainter objects will ultimately contribute to objects missed in Mask R-CNN detection.
  \item For every object in the field, simulate a 512$\times$512 pixel$^2$ \textit{g}-band image with no background using the same catalog from step (i). This second stage is used to produce the object masks. This way, occlusion can be handled from the simulated masks and the network can be trained to identify separate masks for blended sources. Multiple instances of PhoSim are run simultaneously to parallelize this stage.
\end{enumerate}

To speed-up the simulations, we use PhoSim's perfect optics configuration with a simple $\sim$1 arcsec PSF model. This excludes all higher-order optical perturbations and atmospheric effects. Although our simulations are idealized and do not capture all systematics, the neural network only needs to capture the basic morphology of objects and noise in reduced images. Instead, it is sufficient to employ data augmentation afterwards, described in \S\ref{sec:augmentation}, to vary the images and reduce overfitting. Each image has $\sim$150 object masks corresponding to a star or galaxy. We generate 1,000 simulated DECam images in our training set, for a total training set of approximately 150,000 astronomical sources. An example of a typical PhoSim training set image with its galaxy and star masks is shown in Fig.~\ref{fig:example}. We show the distribution of object sizes (in pixels) in our images in Fig.~\ref{fig:size}.

An additional validation set with 250 images is generated using PhoSim in the same manner as the training set.  The validation set is used to reduce overfitting and tune the hyperparameters to find the best model. During the training phase, the model is tested on the validation set to ensure that it is generalizing sufficiently. If the model is solely classifying well on the training set but not the validation set, this is a sign of overfitting. Similarly, we also generate a test dataset of size 50 to get an unbiased evaluation of the performance of the network.

\subsection{Data Standardization}
\label{sec:standardization}

\begin{figure}
    \includegraphics[width=\columnwidth]{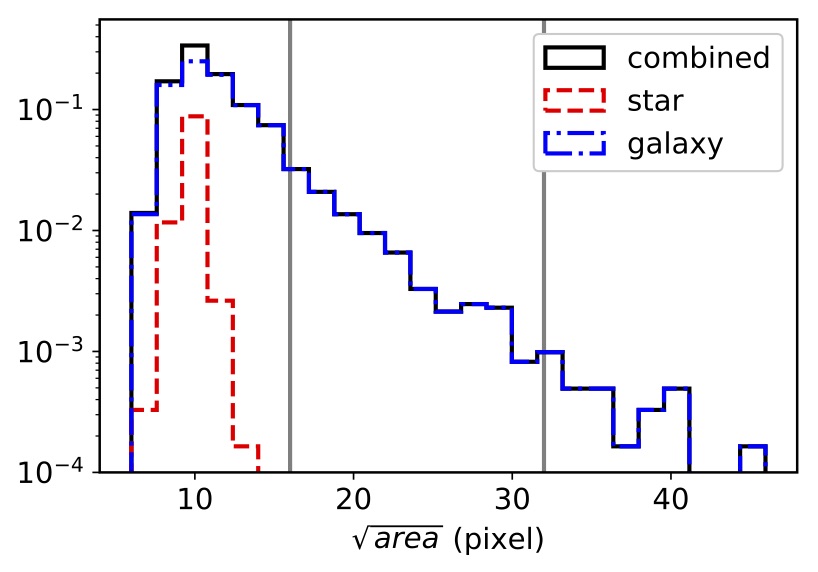}
    \caption{Frequency histogram of bounding box area for objects in the test dataset. The star and galaxy histograms are normalized to the combined stars$+$galaxies distribution. This figure shows the distribution of object sizes in our images. The vertical grey lines show the 16$^2$ and 32$^2$ pixel$^2$ boundaries used for the analysis in section \S\ref{sec:performance}.}
    \label{fig:size}
	\includegraphics[width=\columnwidth]{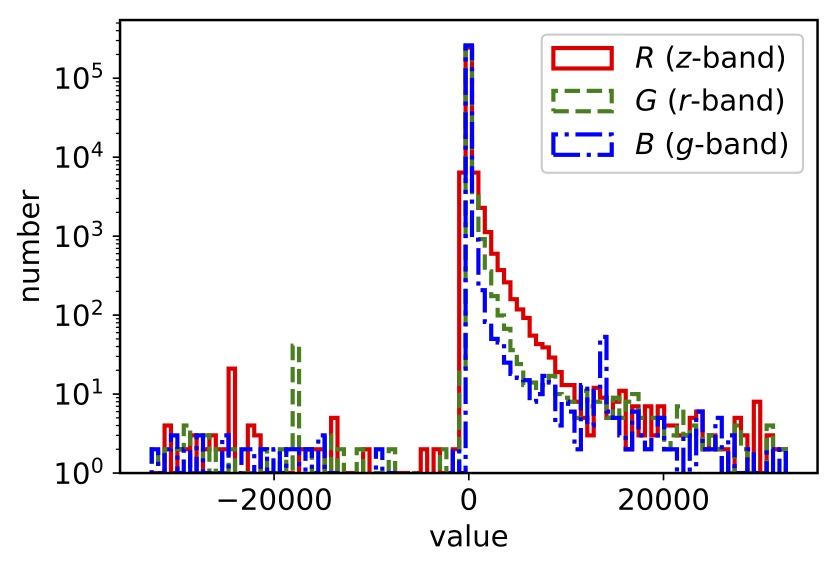}
    \caption{Histogram of normalized pixel values following Eq.~\ref{eq:norm} in a typical simulated training image. The red $R$, green $G$, and blue $B$ channels are shown in their respective colors.}
    \label{fig:norm}
\end{figure}

The color values in each training image are assigned to re-scaled values in each band according to the z-score normalization prescription,

\begin{equation}
    \label{eq:norm}
    \begin{aligned}
    R=A(z-\bar{z})/\sigma_z \\
    G=A(r-\bar{r})/\sigma_r \\
    B=A(g-\bar{g})/\sigma_g \\
    \end{aligned}
\end{equation}

where $R$ is pixel values in the red channel, $\bar{z}$ is the $z$-band mean value, and $\sigma_z$ is the $z$-band standard deviation (and similarly for the green $G$ and blue $B$ channels using the $r$ and $g$ -bands respectively). The scale factor $A$ can be adjusted, for example to correct for exposure times or changes in the instrument sensitivity. We fix $A=10^{4}$ for the training. It is important to preserve the relative values between color channels so that the source color information is retained. We perform this standardization on each set of images (an example is shown in Fig.~\ref{fig:norm}), because the distribution of values can vary greatly in astronomical images depending on objects in the field and observing setup/conditions. This standardization of the image values means our network is not sensitive to the exposure time, gain of the detector, or the final normalization of the reduced images from an image reduction pipeline. This same data standardization is performed on real images during inferencing.

\subsection{Data Augmentation}
\label{sec:augmentation}

We employ the technique of data augmentation \citep{Krizhevsky2012,Dieleman2015} to reduce network overfitting. We perform several image transformations to increase the robustness of our network. These data augmentations preserve object masks and classes. In this work, we augment the data by randomly applying zero to four of the following augmentations:

\begin{itemize}
  \item Rotate: The image and masks are rotated 90, 180, or 270 degrees.
  \item Mirror: The image and masks arrays are mirrored left-right or up-down.
  \item Blur: Smooth the image using a two-dimensional Gaaussian kernel with size $\sigma$ (in pixels) sampled from a random uniform distribution in the interval [$2.0$, $6.0$). This blurs the image and masks, mimicking different PSF sizes.
  \item Add: Add or subtract random values element-wise to each image channel. The possible values are restricted to a range of $\pm10\%$ times the maximum value in the image.
\end{itemize}

These simple augmentations mimic additional observing setups and conditions at little computational expense. With the addition to the random stars and galaxies in the PhoSim images, our network learns rotational invariance. These image augmentations help our network generalize its results to real images or images with slightly different features than the training set.

\subsection{Transfer Learning}

\begin{figure}
	\includegraphics[width=\columnwidth]{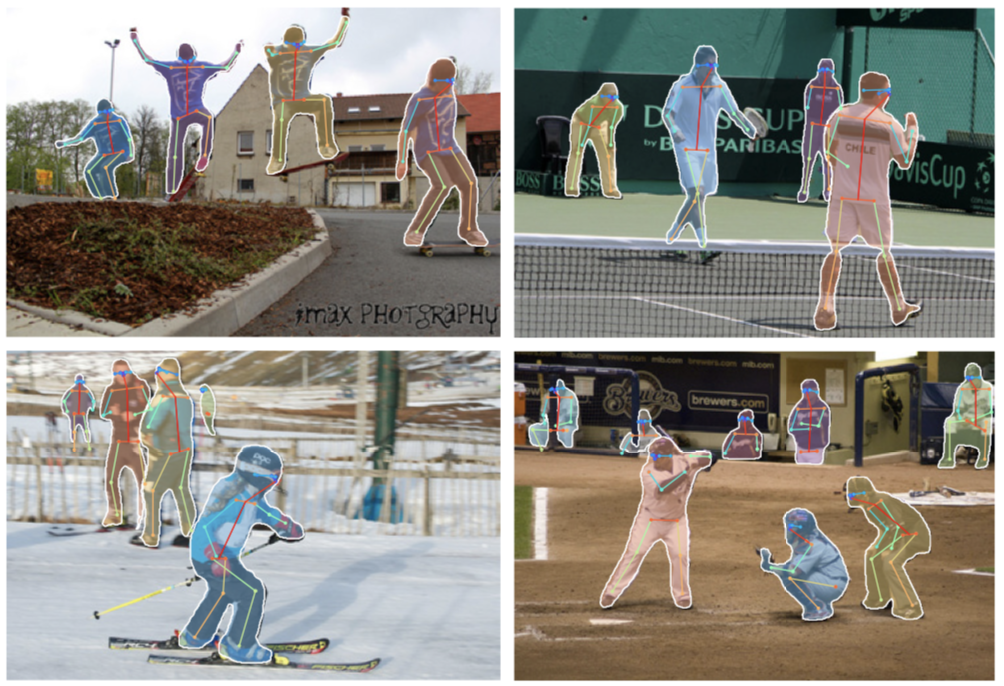}
    \caption{Example of Mask R-CNN person segmentation masks from the MS COCO database (images taken from \citet{He17}, Fig. 7). We employ transfer learning by initializing our training procedure with MS COCO precomputed weights before retraining on simulated astronomical images.}
    \label{fig:mrcnn_coco}
\end{figure}

Transfer learning is a technique in machine learning where networks can generalize knowledge of one task to complete a different but related task. (See \cite{Tan2018} for an overview of deep transfer learning.) In one example of transfer learning, pre-trained weights from one dataset are used as initial conditions for training on another dataset. This improves the speed of training and reduce overfitting of the network. We use Mask R-CNN weights provided by \cite{Matterport17} trained on the Microsoft Common Objects in Context (MS COCO) dataset \citep{Lin2014} as the starting point for our training procedure. MS COCO is a dataset of $\sim$328,000 images with 91 classes of everyday objects (Fig.~\ref{fig:mrcnn_coco}).

\subsection{Training}
\label{sec:training}

Our network is trained in two stages using stochastic gradient descent \citep{Kingma2014}. Stochastic gradient descent updates the model parameters (weights) $\theta_j$ by minimizing the cost function $J(\theta)$ in the equation
\begin{equation}
    \theta_{j+1} = \theta_{j} - \alpha \frac{\partial}{\partial \theta_{j}} J(\theta_j),
\end{equation}
where $\alpha$ is the learning rate. The learning rate is a hyperparameter that is fine-tuned so that the model avoids trapping in local minima and achieves convergence. The first stage performs a re-training of the head layers with a learning rate of $\alpha=10^{-3}$. The second stage trains all layers with a learning rate starting at $\alpha=10^{-4}$ which we decrease progressively to $\alpha=10^{-6}$. We use 50 learning epochs in total (see Appendix~\ref{sec:loss}). Nearly all 1,000 sets of training images and 250 validation images are processed per epoch. By re-scaling our training data to 16-bit integer arrays, we can read-in several sets of images and masks per GPU simultaneously using ResNet-101.

When training, each sampled RoI has an associated multi-task loss, following the form $L = L_\text{cls} + L_\text{box} + L_\text{mask}$ \citep{He17}. Here, $L_\text{cls}$ is the classification loss $-\log p_u$ for ground-truth class $u$ and discrete probability distribution per RoI $p=(p_0,...,p_{K})$ over $K+1$ classes \citep{Girshick2015}. $L_\text{box}$ is the bounding-box loss as defined in \cite{Girshick2015}. Because the mask branch contains $K$ binary masks of resolution $m \times m$ for each of the $K$ classes, a per-pixel sigmoid is applied and thus defines $L_\text{mask}$ as the average binary cross-entropy loss. This specific form of the loss was chosen by \cite{He17} instead of the more commonly used per-pixel softmax and multinomial cross-entropy in fully convolutional network implementation to allow the network to generate masks across classes without competition between them. This is vital for decoupling class and mask prediction, a key feature of the Mask R-CNN architecture.

We use four state-of-the-art NVIDIA Tesla V100 GPUs (each with 5,120 cores and 16 GB high bandwidth memory) to train on 1,000 simulated color images. To speed-up the training, we employ model-based parallelism. Our model is copied onto each GPU where the training workload is divided equally before the weights are brought back together. Our training took $\sim$3 hours to complete (wall time) and reached a total training set losses of $L_\text{cls}=0.209$ $L_\text{box}=0.208$ $L_\text{mask}=0.311$. After this initial cost to train the network, detection and inference can be performed on images in less than a second. Our loss curve during training is shown in Appendix~\ref{sec:loss}.

\section{Results}
\label{sec:results}

To validate our trained Mask R-CNN network, we test its performance against simulated PhoSim images from the test dataset. The test dataset is not used in the training, thus it can be used to give an unbiased estimate of the network performance. Then, we assess our network using a real DECam image of a crowded field and present examples of deblending.

\begin{figure}
	\includegraphics[width=\columnwidth]{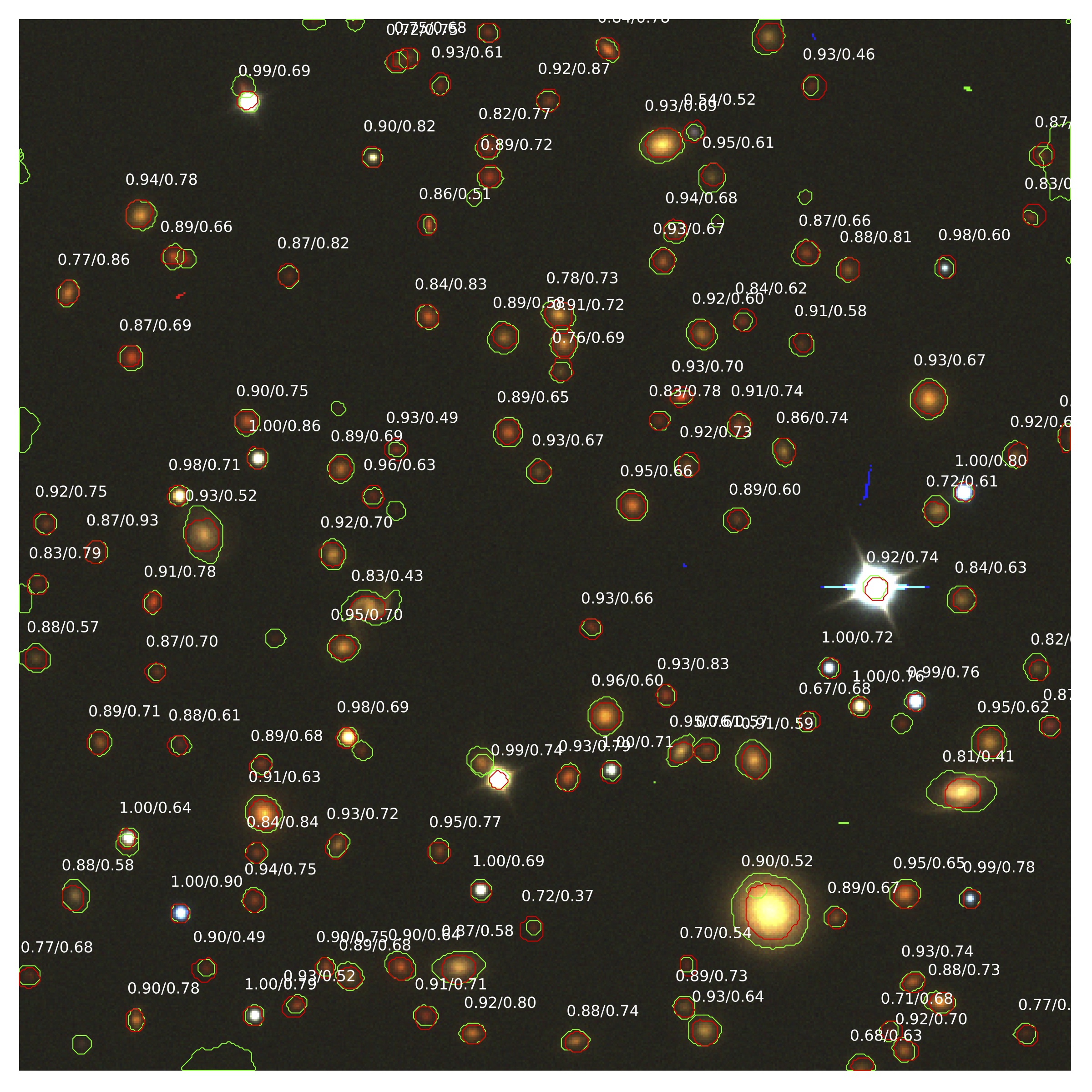}
    \caption{Typical example of detection inference (red) and ground truth (green) in the PhoSim validation dataset. The intersection over union is shown in white text near each positive detection. Note the successful recognition of the saturated star and several blended objects.}
    \label{fig:difference}
\end{figure}

\subsection{Network Performance}
\label{sec:performance}

We use our simulated PhoSim catalog as a truth catalog to validate and evaluate the performance of our Mask R-CNN implementation. Throughout this section we use the test dataset, on which the network is not trained. Thus, we can avoid any misclassification bias that would come from using a real image and catalog. However, there may be systematic differences between the simulated and real images that are not taken into account in this section.

To quantify the performance of our network's classification capability, we calculate the precision and recall for each image in the test dataset. The precision $p$ (purity) and recall $r$ (completeness) are defined as follows:

\begin{equation}
    p=\frac{\text{TP}}{\text{TP}+\text{FP}}
\end{equation}

\begin{equation}
    r=\frac{\text{TP}}{\text{TP}+\text{FN}}
\end{equation}

\begin{figure}
	\includegraphics[width=\columnwidth]{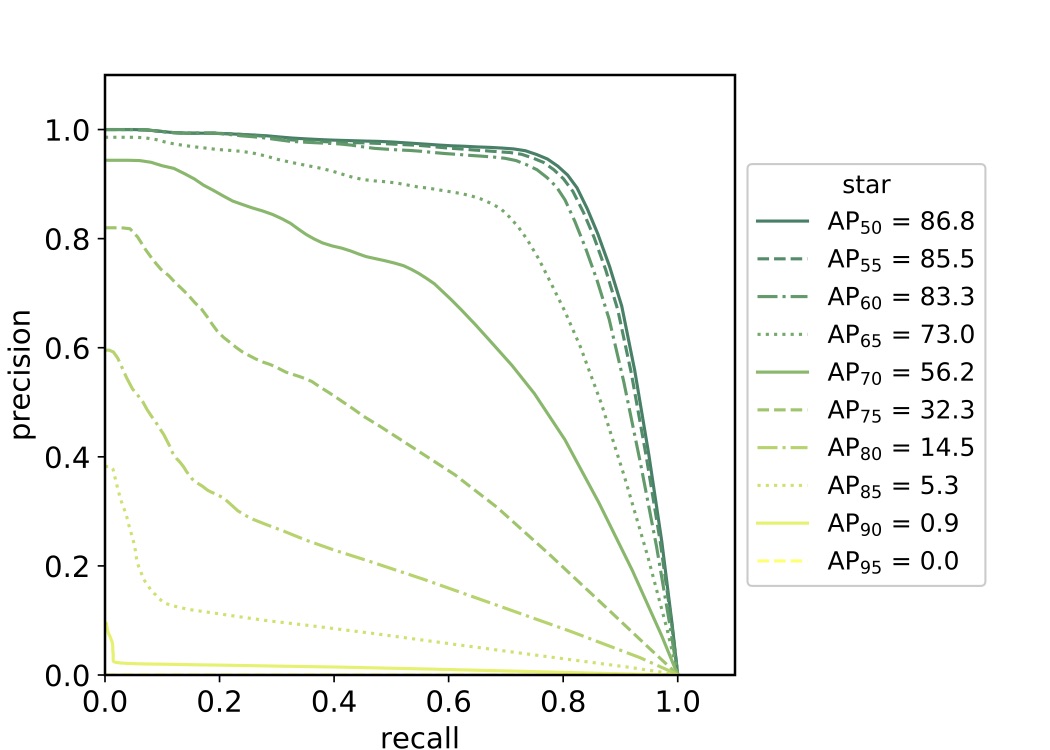}
    \caption{Precision-recall curves and mean AP scores calculated at varying IOU thresholds averaged over the test dataset for stars. We find IOU=0.5 gives a mean AP score of 86.8.}
    \label{fig:precision-recall1}
	\includegraphics[width=\columnwidth]{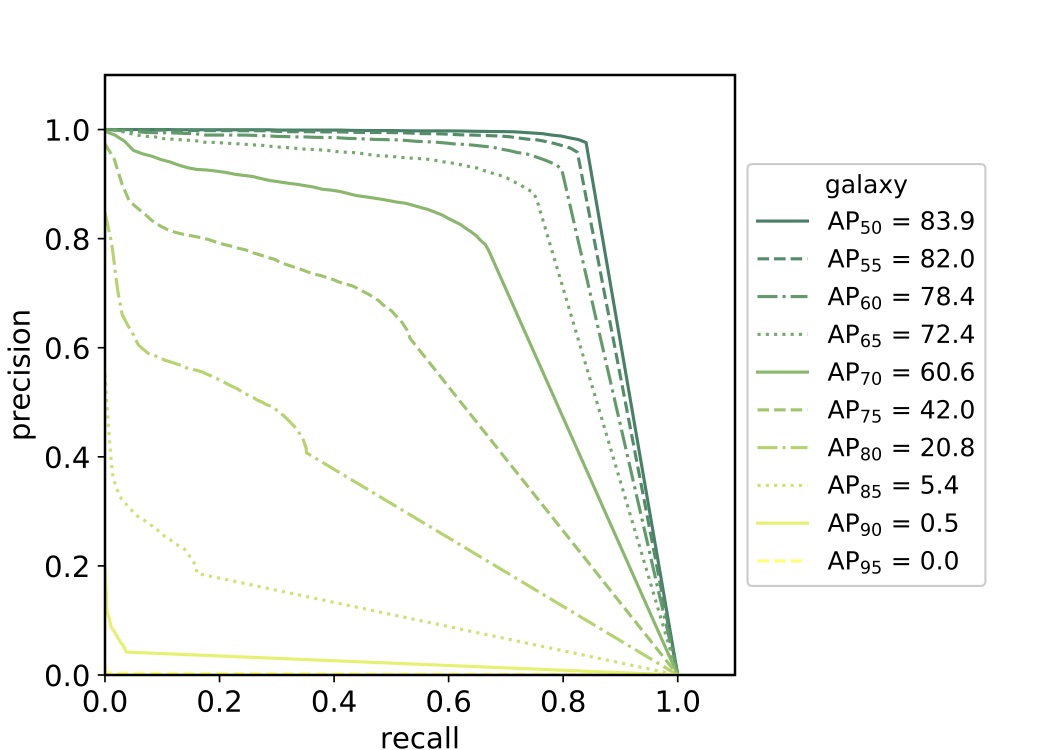}
    \caption{Precision-recall curves and mean AP scores calculated at varying IOU thresholds averaged over the test dataset for galaxies. We find IOU=0.5 gives a mean AP score of 83.9.}
    \label{fig:precision-recall2}
    	\includegraphics[width=\columnwidth]{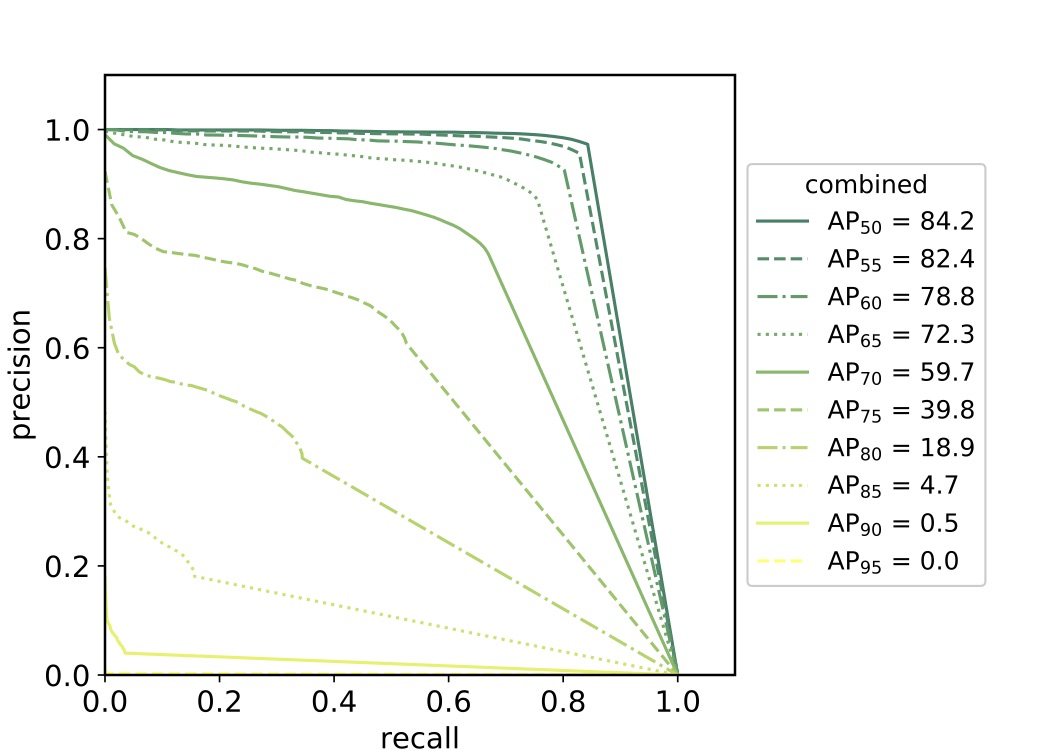}
    \caption{Precision-recall curves and mean AP scores calculated at varying IOU thresholds averaged over the test dataset calculated jointly for both stars and galaxies. We find IOU=0.5 gives a mean AP score of 84.2.}
    \label{fig:precision-recall3}
\end{figure}

where TP$=$true positive, TN$=$true negative, FP$=$false positive, FN$=$false negative. A detection is considered positive if its ranked output (detection confidence) is greater than a given minimum detection confidence threshold (this value can be adjusted, but we choose 0.5 throughout this work). Note that the detection confidence is not the same as the Bayesian significance given by a $S/N$ calculation. By comparing our network's classification results to the PhoSim catalog, we can evaluate the precision at various recalls. In this analysis, we analyze the precision and recall for both star and galaxy classes as well as both classes jointly (combined).

We adopt a metric commonly used in the machine learning community called the average precision (AP) score\footnote{The AP score has largely superseded the older area under the receiver operating characteristic (ROC) curve in the computer vision community in favor of the AP score's greater sensitivity for high-performing networks.} \citep[see][]{Everingham2010}. The AP score is simply the area under the precision-recall curve, averaged for each object in an image at discrete recall levels:

\begin{equation}
    \text{AP}=\frac{1}{51}\sum_{r\in\{0,0.02,...,1.0\}} p(r)
\end{equation}

where $p(r)$ is maximum the precision in bin $\Delta r$. We average the precision-recall curves and mean AP scores for all 50 images in the test dataset. This procedure is then done for the intersection over union (IOU) thresholds $\text{IOU}\in\{0.5,0.55,\ldots,0.95\}$. The IOU is defined as the area of the intersection of the predicted and ground truth masks divided by the area of the union of the predicted and ground truth masks,

\begin{equation}
    \text{IOU}=\frac{\text{\emph{area}}(\text{mask}_\text{predicted} \cap \text{mask}_\text{truth})}{\text{\emph{area}}(\text{mask}_\text{predicted} \cup \text{mask}_\text{truth})}.
\end{equation}

An RoI detection is considered positive if its IOU is greater than the IOU threshold. We therefore expect fewer positive detections at greater IOU thresholds. We show an example of detection masks overlayed with ground truth masks in Fig.~\ref{fig:difference} in a simulated test image. The results shown in Figs.~\ref{fig:precision-recall1}, \ref{fig:precision-recall2}, and \ref{fig:precision-recall3} summarize the performance of our network against PhoSim ground truth images in the test dataset. We also show the confusion matrix at IOU thresholds of 0.5 (Fig.~\ref{tab:confusion1}) and 0.75 (Fig.~\ref{tab:confusion1}).

In Table~\ref{tab:performance}, we calculate various AP metrics to evaluate the performance of our work for stars, galaxies, and a joint star$+$galaxy (combined) evaluation. We use the standard AP score variants used in the computer vision community defined as follows:
\begin{itemize}
  \item AP: AP score averaged over IOU thresholds of 0.5 to 0.95.
  \item AP$_{50}$: AP score at IOU threshold of 0.5 (50\%).
  \item AP$_{75}$: AP score at IOU threshold of 0.75 (75\%).
  \item AP$_{\rm S}$: AP score of small sized ($\text{\emph{area}}<16^2$ pixel$^2$) objects bounding box at IOU threshold of 0.5.
  \item AP$_{\rm M}$: AP score of medium sized ($16^2{<}\text{\emph{area}}<32^2$ pixel$^2$) objects bounding box at IOU threshold of 0.5.
  \item AP$_{\rm L}$: AP score of large sized ($\text{\emph{area}}>32^2$ pixel$^2$) objects at IOU bounding box threshold of 0.5.
\end{itemize}
These metrics show how our Mask R-CNN implementation performs across mask accuracy (by varying the IOU threshold) and scales (by evaluating for different bounding box sizes), as well as facilitates comparison to other works. Note, that our definitions of AP$_{\rm S}$, AP$_{\rm M}$, and AP$_{\rm L}$ differ from those used by the MS COCO evaluation, because astronomical objects are generally much smaller in area in pixels.

\begin{table}
\centering
\begin{tabular}{lcccccc}
  \hline
  \hline
 Class & AP & AP$_{50}$ & AP$_{75}$ & AP$_{\rm S}$ & AP$_{\rm M}$ & AP$_{\rm L}$ \\
 \hline
 star & 48.6 & 86.8 & 32.3 & 86.8 & -- & -- \\
 galaxy & 49.6 & 83.9 & 42.0 & 75.3 & 7.6 & 0.3 \\
 combined & 49.0 & 84.2 & 39.8 & 76.4 & 7.0 & 0.3\\
\hline
\end{tabular}
\caption{Summary table of our AP score metrics calculated on the test dataset. Note their definitions in the text. Although our network performs well for small sources, its performance rapidly decreases for larger sources. This is likely due to the scarcity of sources with \emph{area} > 16$^2$ pixel$^2$ in the training images. All stars have bounding box sizes \emph{area} < 16$^2$ pixel$^2$.
\label{tab:performance}
}
\end{table}

\begin{table}
\centering
\begin{tabular}{cc|cc}
\multicolumn{1}{c}{} &\multicolumn{1}{c}{} &\multicolumn{2}{c}{Predicted} \\
\multicolumn{1}{c}{} & \multicolumn{1}{c}{} & \multicolumn{1}{c}{star} &
\multicolumn{1}{c}{galaxy} \\ \hline
\multirow{1}{*}{\rotatebox{90}{Truth}}
& star  & 585 & 39   \\
& galaxy  & 78   & 6302 \\ \hline
\end{tabular}
\caption{Confusion matrix calculated on the test dataset for stars/galaxies at an IOU threshold of 0.5.
\label{tab:confusion1}
}
\end{table}

\begin{table}
\centering
\begin{tabular}{cc|cc}
\multicolumn{1}{c}{} &\multicolumn{1}{c}{} &\multicolumn{2}{c}{Predicted} \\
\multicolumn{1}{c}{} & \multicolumn{1}{c}{} & \multicolumn{1}{c}{star} &
\multicolumn{1}{c}{galaxy} \\ \hline
\multirow{1}{*}{\rotatebox{90}{Truth}}
& star  & 330 & 68   \\
& galaxy  & 333   & 6273 \\ \hline
\end{tabular}
\caption{Confusion matrix calculated on the test dataset for stars/galaxies at an IOU threshold of 0.75.
\label{tab:confusion2}
}
\end{table}

\subsection{Deblending}

\begin{figure}
	\includegraphics[width=\columnwidth]{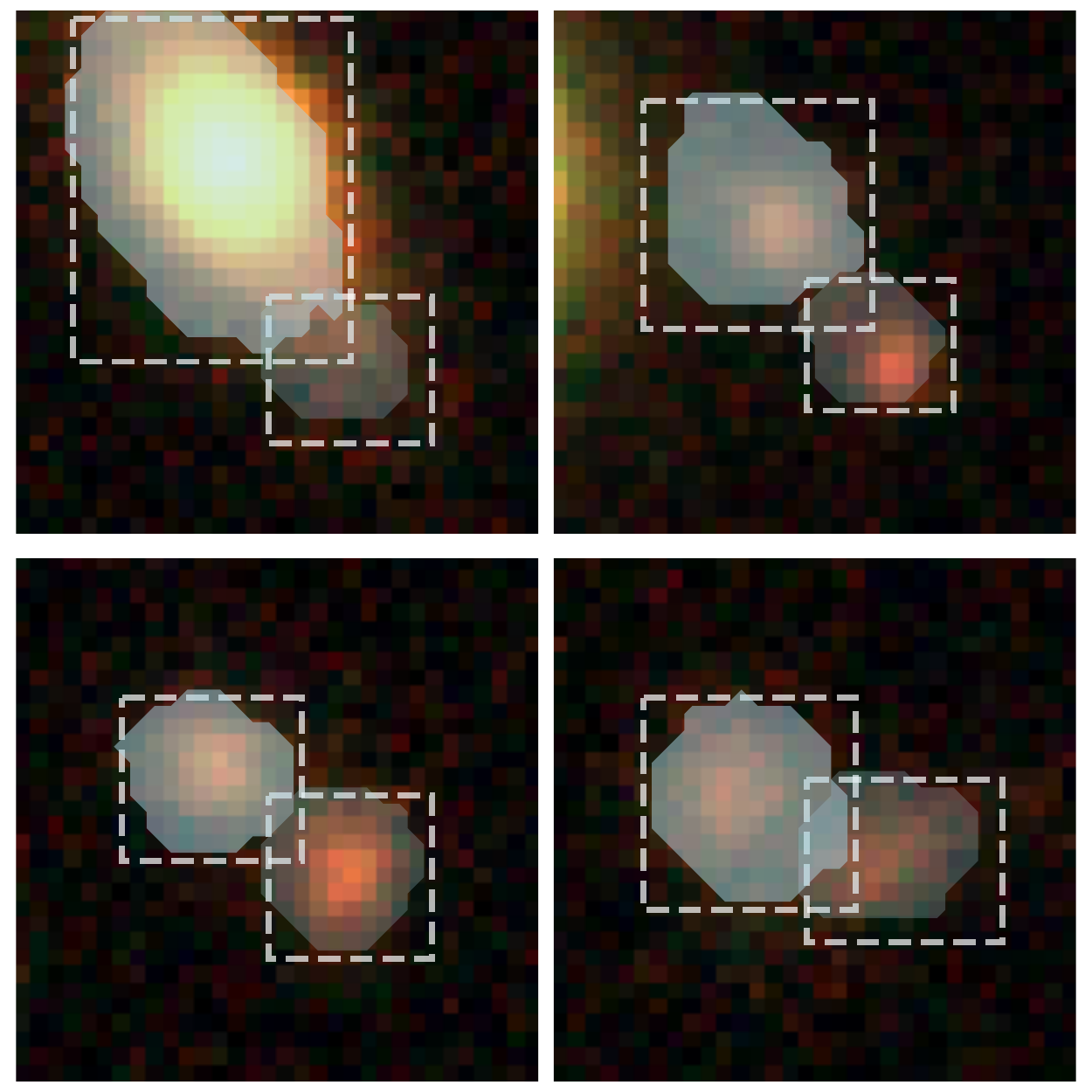}
    \caption{Examples of close blends identified as separate galaxies in a real DECaLs image using our deep learning technique. Bounding box edges are shown as white dashed lines, and masks are shown as transparent white regions. An additional $S/N$ or detection confidence cutoff may be applied afterwards to remove false or spurious detections.}
    \label{fig:blend}
\end{figure}

\begin{figure*}
	\includegraphics[width=\textwidth]{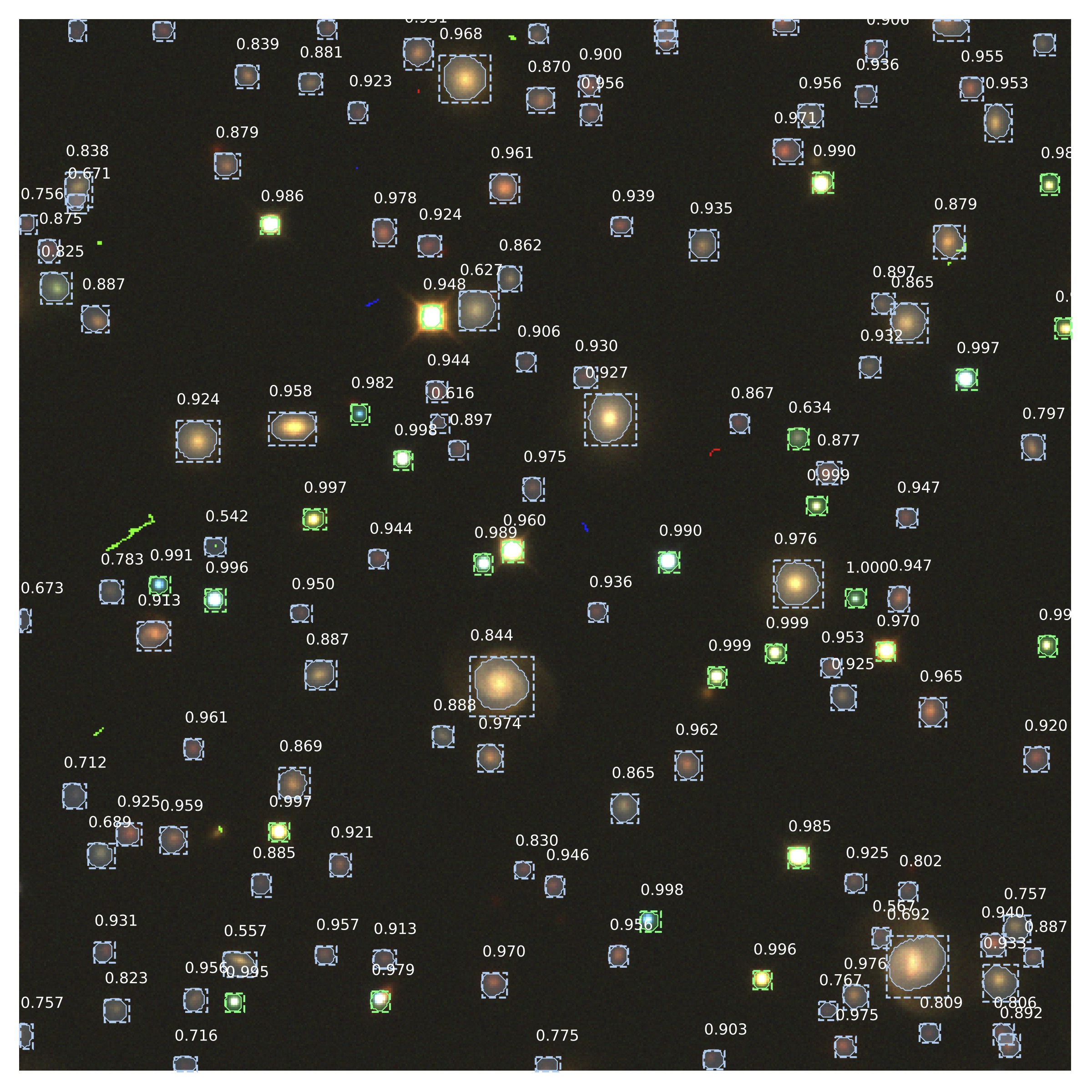}
    \caption{Typical example of detection inference in a full test dataset simulated image. Although this image is not used in the training, it is a typical example of what the training images look like (generated using the same PhoSim settings and procedure). Galaxy masks are shown in light blue and star masks are shown in green. The detection confidence the object belongs to that particular class is shown above each mask. Note several blends are successfully identified as separate objects, particularly at the lower-right.}
    \label{fig:inssim}
\end{figure*}

\begin{figure*}
	\includegraphics[width=\textwidth]{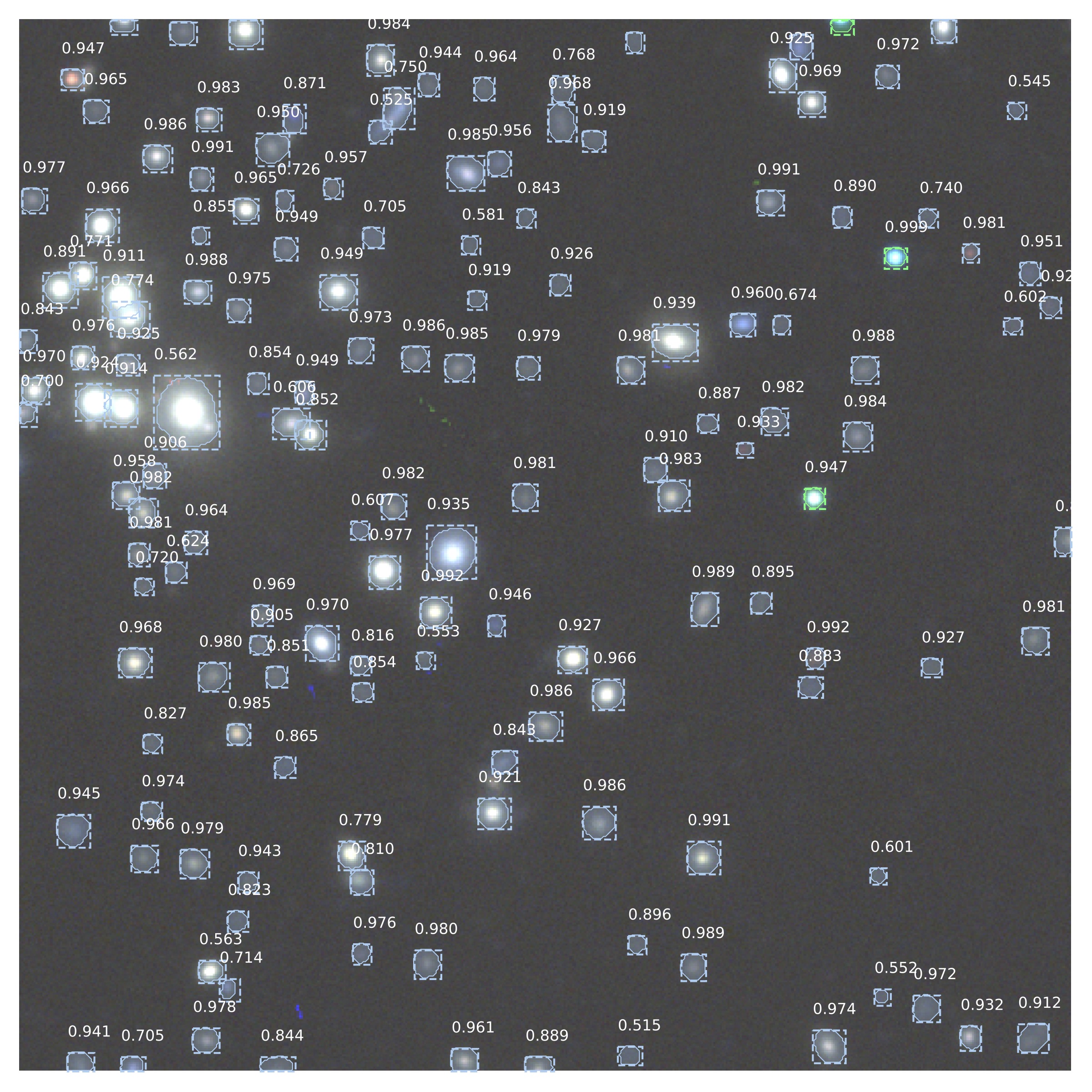}
    \caption{Example of detection inference in a real DECaLS image of ACO 1689. Galaxy masks are shown in light blue and star masks are shown in green. The detection confidence the object belongs to that particular class is shown above each mask. The central nucleus of the cluster is located in the upper-left of the image.}
    \label{fig:insreal}
\end{figure*}

To test the deblending capability of our code, we simulate a PhoSim image with a high number density of galaxies to mimic the DECaLS images. One relevant inference configuration parameter is $\texttt{DETECTION\_NMS\_THRESHOLD}$. This parameter sets the intersection threshold for non-maximum suppression (NMS). Mask R-CNN and other sliding window -based approaches typically result in several high-confidence detections for individual objects. NMS rejects low-confidence bounding boxes that have high IOU overlaps with another bounding box. We use an IOU threshold of 0.3 for typical images, although this threshold may be lowered in very dense regions, such as the dense central nucleus of a cluster of galaxies, to increase the likelihood that close blends are identified as distinct objects.

In Fig.~\ref{fig:blend}, we demonstrate several examples of deblending performed using our Mask R-CNN technique. The network recognizes close blends and generates segmentation maps that encompass each object. Additional post-processing may be done in some scenarios, such as a $S/N$ cutoff or a smoothing refinement of the masks. Presently, we allow some pixels to overlap with multiple objects after NMS is applied. Ultimately, the pixels that lie inside each masked region can simply be used to isolate objects.

\subsection{Inference on real images}

To investigate the performance of our network on real images, we use the public data release 7 DECaLs coadds available on the NOAO science archive\footnote{\hyperlink{http://archive.noao.edu}{http://archive.noao.edu}}. We wish to assess our network in the limit of a crowded extragalactic field. Therefore, we use images centered on clusters of galaxies in the Abell Catalog \citep{Olowin1988,Abell1989}. Specifically, we use DECaLS images of ACO 1689. ACO 1689 is a rich \citep[class 4;][]{Abell1989} cluster with a type II-III Bautz-Morgan classification \citep{Bautz1970} at $z=0.183$. ACO 1689's richness and the fact that it has been extensively studied \citep[e.g.][]{Taylor1998,Tyson1995} and imaged by DECam makes it a good target for our tests. We show the result of a detection/inference on a simulated image (Fig.~\ref{fig:inssim}) and a real image (Fig.~\ref{fig:insreal}) for comparison.

We note that the PSF appears significantly larger in the real images compared to the simulated images. This may is likely due to choosing idealized simulated images that excludes some atmospheric or optical perturbations that enlarge the PSF. Another contributing factor is the DECaLS images are coadds taken in varying seeing conditions, which may results in a larger PSF. Despite this, our neural network is not overly-sensitive to the PSF size because we employ a Gaussian blur augmentation. However, very bright stars which have significant blooming can result in a badly-fitting mask or misclassification as a galaxy and a low IOU score for that detection.

\subsection{Ultra-Deep Fields}

\begin{figure*}
	\includegraphics[width=\textwidth]{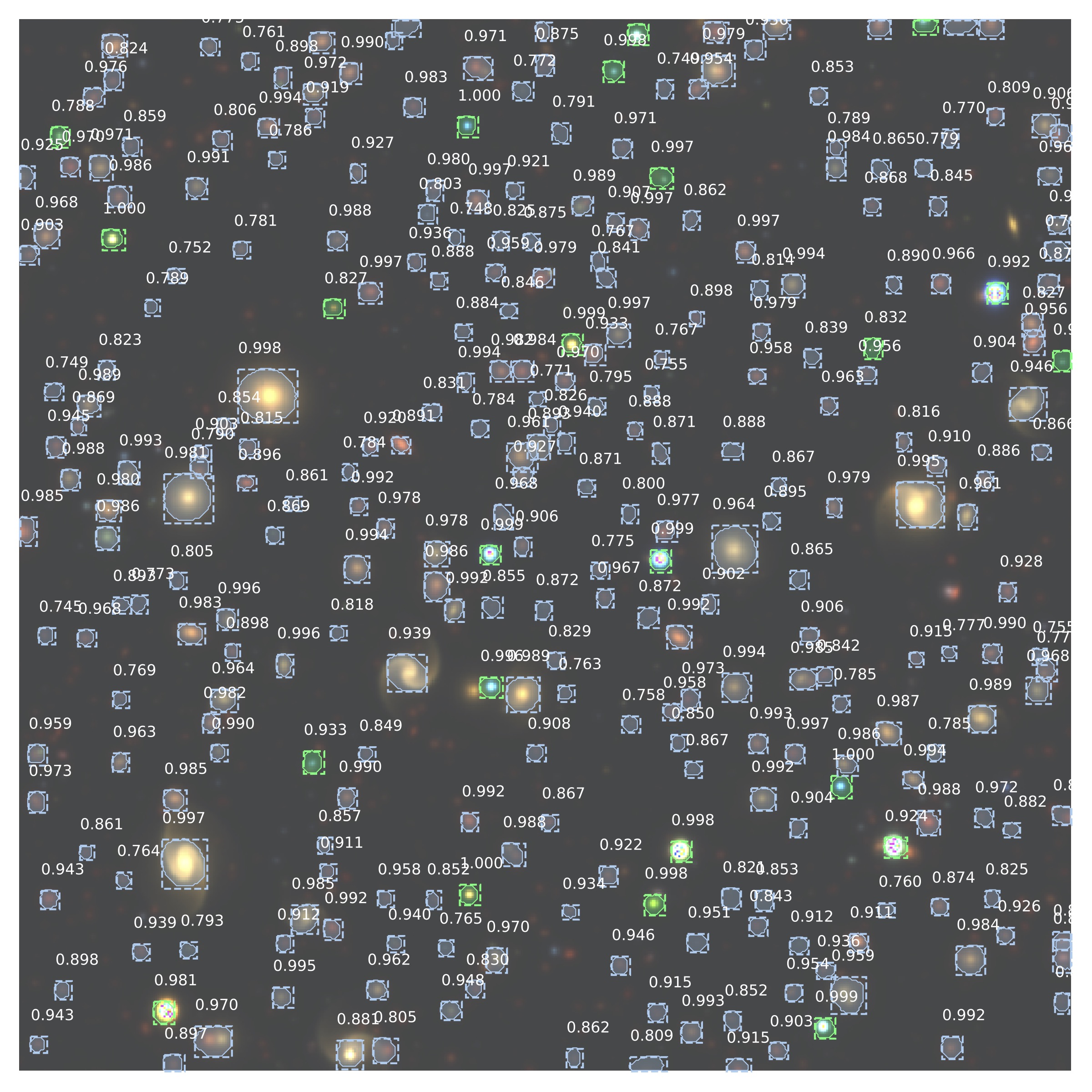}
    \caption{An example of an ultra-deep 30 hour simulation and detection inference. Galaxy masks are shown in light blue and star masks are shown in green. The detection confidence the object belongs to that particular class is shown above each mask. Many of the fainter and higher redshift galaxies are not detected with a confidence threshold of 0.5 or higher because they appear redder than the galaxies in our training data. However, nearly all of more typical sources are detected.}
    \label{fig:inssimlong}
\end{figure*}

We investigate how the richness (source density) of the field affects our networks performance. In general, a deeper and more crowded field is likely to contain a larger number overlapping objects. Following the procedure outlined in \S\ref{sec:trainingset}, we simulate additional images with PhoSim using an exposure time of 30 hours with galaxies as faint as $g=28$. This deep field mimicks the depth of the Subrau HSC ultra-deep field or ten-year LSST coadds, containing $\sim750$ galaxies/arcmin$^2$. For these simulations, we turn of charge saturation and cosmic rays in PhoSim. The result is shown in (Fig.~\ref{fig:inssimlong}). We then calculate the mean AP score in this image at an IOU threshold of 0.5, following the same procedure outlined in \S\ref{sec:performance}. We find $\text{AP}=52$ (stars) and $\text{AP}=7.6$ (galaxies). Our network is not sensitive to the number of density of sources in the image, because the number of objects detected increases in proportion to the number in the simulations. However, galaxies with redshifted SEDs that are too faint to appear in the training images are not detected. We could mitigate this effect and increase the AP score by training on images with longer exposure times with higher-redshift galaxies or allowing a lower detection confidence threshold.

\section{Discussion}
\label{sec:discussion}

One of the main strengths of this technique is its speed and ease of use after training. After training is completed, we find individual images can be processed in 100 milliseconds or less depending on the GPU(s) used (not including time to read/write the results). This and similar machine learning techniques have this important strength, which is particularly relevant for large astronomical surveys. In addition, our Mask R-CNN package performs all tasks of object detection, classification, and segmentation in parallel in a simple conceptual framework. Finally, our package is insensitive to configuration parameters, such as the density of sources in the field, PSF size, and exposure time. Instead, a large and diverse set of training images and additional data augmentations is needed.

Here we describe briefly the new and unique contributions of this work:

\begin{itemize}
    \item Establish a novel technique for instance segmentation (masking) in astronomical images.
    \item Show how existing deep learning techniques in the field of computer vision can be used to solve difficult problems such as deblending in astronomy.
    \item Use of multi-band information to perform star/galaxy classification and deblending.
    \item Train a neural network on large, realistic simulated images approaching the size of whole CCDs.
\end{itemize}

There are several limitations to the machine learning techniques used in this work. First, it can be hard to correct for the systematic differences between simulated and real images. Issues arise if the real images do not mimic the simulated images. This can occur if the data in one band is missing or noisy (confusing the colors of objects). However, transfer learning could be used to perform a fine-tuning re-training for these scenarios. Occasionally, the neural network will confuse bright galaxy bulges for stars if the network has not been trained to a sufficient loss. This can be solved by training more extensively, employing non-maximum suppression to quickly reject these stars, or perhaps using simulated galaxies more brighter or more realistic bulges. Finally, it can be challenging to generate a realistic training set that accounts for the vast differences in images in different regions of the sky. For example, globular clusters, near the galactic center of the Milky Way, or images of large/extended galaxies may be challenging regimes. However, we expect that these problems can be mitigated if these types of objects were added to the simulated training data. Although care must be taken when simulating images of these ultra-dense regions. Because PhoSim works by generating a catalog from a three-dimensional space, we must not train for detections on sources that are completely obscured by another object.

In the future, we will explore masking additional features such as cosmic rays and bleed trails. These distinct features should be straightforward for Mask R-CNN to learn. We may also explore different network architectures for instance segmentation, such as the popular U-Net \citep{Ronneberger2015}. U-Net may be better-suited to full-scale astronomical images because it does not create regions around each object, which can lead to inefficiencies in the training. The down-side is that U-Net can only inherently perform semantic segmentation (i.e., not distinguishing masks of individual objects of the same class). Therefore, these methods must eventually fall-back on traditional techniques to handle mask occlusion, and may suffer from the same issues current pipelines face in crowded fields.

Moreover, we will explore (a) executing and tiling results on a full-scale dataset with images taken under a variety of conditions, such as DES or HSC deep coadd images; (b) predicting probability contours for objects by training using different mask sizes or IOU cutoffs, add photometric redshift prediction, adding additional galaxy classes in place of the generic ``galaxy" classification, such as spiral, elliptical, irregular, and lensed; (c) implementing a more robust interface for training and configurations. Mask R-CNN can also be used on video for real-time instance segmentation, which may have interesting applications for LSST and time-domain astronomy.

There may also be additional ways to optimize our Mask R-CNN implementation (both training and detection) on astronomical image data. For example, it is possible that different contrast stretching such as \cite{Lupton2004} \citep[see][]{Gonzalez2018} or backbone architecture will produce better results on astronomical images. The standard techniques adopted in this work (linear scaling with the ResNet-101 backbone) used on terrestrial images in the MS COCO dataset may not be as optimal for astronomical data which typically have smaller objects at lower $S/N$. Based on our evaluation, using training images which are \emph{less} realistic, but include a larger variety of objects and with a more uniform distribution of sizes may be appropriate.

Because our network works well on real images, this work also acts as a unique and real-world validation of PhoSim itself. Several additional telescopes/instruments are implemented in PhoSim, including LSST and Subaru HSC. Therefore, extensions of this work to other telescopes would be straightforward. We would welcome further extensions of the \texttt{Astro R-CNN} code repository \citep{repo} to improve the training, performance, and usability. Transfer learning from this work to other telescopes or surveys could also be used to shorten training time and with little required GPU resources.

\section{Conclusions}
\label{sec:conclusions}

In this work, we develop a new deep-learning method for classifying and deblending sources in astronomical images. Using PhoSim, we generate simulated images and catalogs used as a ground truth comparison for supervised machine learning. Our code, $\texttt{Astro R-CNN}$ efficiently performs all tasks of source detection, classification, and deblending in one package. The network is robust to object overlaps and mask occlusion, resulting in clean deblends for significantly blended sources.

We evaluate the performance of our network using the AP score metric, and show the precision and recall curves for stars and galaxies. Our network performs well at moderate IOU thresholds. We measure a precision of 92\% at 80\% recall for stars and a precision of 98\% at 80\% recall for galaxies in a typical field with $30$ galaxies/arcmin$^2$ with an IOU threshold of 0.5 and a minimum detection confidence threshold of 0.5. We show multiple examples of deblending in real and simulated images, including in the central nucleus of an Abell cluster of galaxies. Because of the region-based nature of the Mask R-CNN neural network, the results are insensitive to the density of sources in the image therefore may be used at different galactic latitudes. If care is taken to configure and train $\texttt{Astro R-CNN}$ properly, it may be used on a variety of real images. The depth and quantity of images in current and future deep optical surveys demands efficient and robust techniques such as this one. We suggest future endeavors use this work as an example to apply even newer and rapidly-advancing techniques from the computer vision community to astronomy.

\section*{Acknowledgements}

This work utilizes resources supported by the National Science Foundation's Major Research Instrumentation program, grant \#1725729, as well as the University of Illinois at Urbana-Champaign. We thank Dr. Volodymyr Kindratenko and Dr. Dawei Mu at the National Center for Supercomputing Applications for their assistance with the GPU cluster used in this work. This work is based on projects started during the 2019 graduate-level AI seminar at the Department of Astronomy, University of Illinois at Urbana-Champaign. We thank the anonymous referees for helpful comments.

We acknowledge use of Matplotlib \citep{Hunter2007}, a community-developed Python library for plotting. This research made use of Astropy,\footnote{\hyperlink{http://www.astropy.org}{http://www.astropy.org}} a community-developed core Python package for Astronomy \citep{astropy:2013, astropy:2018}. This research has made use of NASA's Astrophysics Data System.

NOAO is operated by the Association of Universities for Research in Astronomy (AURA) under a cooperative agreement with the National Science Foundation.

This project used data obtained with the Dark Energy Camera (DECam), which was constructed by the Dark Energy Survey (DES) collaboration. Funding for the DES Projects has been provided by the U.S. Department of Energy, the U.S. National Science Foundation, the Ministry of Science and Education of Spain, the Science and Technology Facilities Council of the United Kingdom, the Higher Education Funding Council for England, the National Center for Supercomputing Applications at the University of Illinois at Urbana-Champaign, the Kavli Institute of Cosmological Physics at the University of Chicago, Center for Cosmology and Astro-Particle Physics at the Ohio State University, the Mitchell Institute for Fundamental Physics and Astronomy at Texas A\&M University, Financiadora de Estudos e Projetos, Fundacao Carlos Chagas Filho de Amparo, Financiadora de Estudos e Projetos, Fundacao Carlos Chagas Filho de Amparo a Pesquisa do Estado do Rio de Janeiro, Conselho Nacional de Desenvolvimento Cientifico e Tecnologico and the Ministerio da Ciencia, Tecnologia e Inovacao, the Deutsche Forschungsgemeinschaft and the Collaborating Institutions in the Dark Energy Survey. The Collaborating Institutions are Argonne National Laboratory, the University of California at Santa Cruz, the University of Cambridge, Centro de Investigaciones Energeticas, Medioambientales y Tecnologicas-Madrid, the University of Chicago, University College London, the DES-Brazil Consortium, the University of Edinburgh, the Eidgenossische Technische Hochschule (ETH) Zurich, Fermi National Accelerator Laboratory, the University of Illinois at Urbana-Champaign, the Institut de Ciencies de l'Espai (IEEC/CSIC), the Institut de Fisica d'Altes Energies, Lawrence Berkeley National Laboratory, the Ludwig-Maximilians Universitat Munchen and the associated Excellence Cluster Universe, the University of Michigan, the National Optical Astronomy Observatory, the University of Nottingham, the Ohio State University, the University of Pennsylvania, the University of Portsmouth, SLAC National Accelerator Laboratory, Stanford University, the University of Sussex, and Texas A\&M University.




\bibliographystyle{mnras}
\bibliography{article} 

\begin{thebibliography}{}
\makeatletter
\relax
\def\mn@urlcharsother{\let\do\@makeother \do\$\do\&\do\#\do\^\do\_\do\%\do\~}
\def\mn@doi{\begingroup\mn@urlcharsother \@ifnextchar [ {\mn@doi@}
  {\mn@doi@[]}}
\def\mn@doi@[#1]#2{\def\@tempa{#1}\ifx\@tempa\@empty \href
  {http://dx.doi.org/#2} {doi:#2}\else \href {http://dx.doi.org/#2} {#1}\fi
  \endgroup}
\def\mn@eprint#1#2{\mn@eprint@#1:#2::\@nil}
\def\mn@eprint@arXiv#1{\href {http://arxiv.org/abs/#1} {{\tt arXiv:#1}}}
\def\mn@eprint@dblp#1{\href {http://dblp.uni-trier.de/rec/bibtex/#1.xml}
  {dblp:#1}}
\def\mn@eprint@#1:#2:#3:#4\@nil{\def\@tempa {#1}\def\@tempb {#2}\def\@tempc
  {#3}\ifx \@tempc \@empty \let \@tempc \@tempb \let \@tempb \@tempa \fi \ifx
  \@tempb \@empty \def\@tempb {arXiv}\fi \@ifundefined
  {mn@eprint@\@tempb}{\@tempb:\@tempc}{\expandafter \expandafter \csname
  mn@eprint@\@tempb\endcsname \expandafter{\@tempc}}}

\bibitem[\protect\citeauthoryear{Abadi et~al.,}{Abadi et~al.}{2016}]{Abadi2016}
Abadi M.,  et~al., 2016, in 12th USENIX Symposium on Operating Systems Design
  and Implementation (OSDI 16). pp 265--283

\bibitem[\protect\citeauthoryear{{Abbott} et~al.,}{{Abbott}
  et~al.}{2018}]{Abbott2018}
{Abbott} T.~M.~C.,  et~al., 2018, \mn@doi [\apjs] {10.3847/1538-4365/aae9f0},
  \href {https://ui.adsabs.harvard.edu/abs/2018ApJS..239...18A} {239, 18}

\bibitem[\protect\citeauthoryear{Abdulla}{Abdulla}{2017}]{Matterport17}
Abdulla W.,  2017, Mask R-CNN for object detection and instance segmentation on
  Keras and TensorFlow, \url{https://github.com/matterport/Mask_RCNN}

\bibitem[\protect\citeauthoryear{{Abell}, {Corwin}  \& {Olowin}}{{Abell}
  et~al.}{1989}]{Abell1989}
{Abell} G.~O.,  {Corwin} Jr. H.~G.,   {Olowin} R.~P.,  1989, \mn@doi [\apjs]
  {10.1086/191333}, \href
  {https://ui.adsabs.harvard.edu/abs/1989ApJS...70....1A} {70, 1}

\bibitem[\protect\citeauthoryear{{Aihara} et~al.,}{{Aihara}
  et~al.}{2018}]{Aihara2018}
{Aihara} H.,  et~al., 2018, \mn@doi [\pasj] {10.1093/pasj/psx081}, \href
  {https://ui.adsabs.harvard.edu/abs/2018PASJ...70S...8A} {70, S8}

\bibitem[\protect\citeauthoryear{{Aihara} et~al.,}{{Aihara}
  et~al.}{2019}]{Aihara2019}
{Aihara} H.,  et~al., 2019, arXiv e-prints, \href
  {https://ui.adsabs.harvard.edu/abs/2019arXiv190512221A} {p. arXiv:1905.12221}

\bibitem[\protect\citeauthoryear{{Allende Prieto}, {Barklem}, {Lambert}  \&
  {Cunha}}{{Allende Prieto} et~al.}{2004}]{Prieto2004}
{Allende Prieto} C.,  {Barklem} P.~S.,  {Lambert} D.~L.,   {Cunha} K.,  2004,
  \mn@doi [\aap] {10.1051/0004-6361:20035801}, \href
  {https://ui.adsabs.harvard.edu/abs/2004A&A...420..183A} {420, 183}

\bibitem[\protect\citeauthoryear{{Amiaux} et~al.}{{Amiaux}
  et~al.}{2012}]{Amiaux12}
{Amiaux} J.,  et~al., 2012, Euclid Mission: building of a reference survey,
  \mn@doi{10.1117/12.926513}, \url {https://doi.org/10.1117/12.926513}

\bibitem[\protect\citeauthoryear{{Andreon}, {Gargiulo}, {Longo}, {Tagliaferri}
  \& {Capuano}}{{Andreon} et~al.}{2000}]{Andreon00}
{Andreon} S.,  {Gargiulo} G.,  {Longo} G.,  {Tagliaferri} R.,   {Capuano} N.,
  2000, \mn@doi [\mnras] {10.1046/j.1365-8711.2000.03700.x}, \href
  {https://ui.adsabs.harvard.edu/abs/2000MNRAS.319..700A} {319, 700}

\bibitem[\protect\citeauthoryear{{Arneson}}{{Arneson}}{2013}]{Arneson2013}
{Arneson} R.~A.,  2013, Master's thesis, University of California, Irvine

\bibitem[\protect\citeauthoryear{{Astropy Collaboration} et~al.,}{{Astropy
  Collaboration} et~al.}{2013}]{astropy:2013}
{Astropy Collaboration} et~al., 2013, \mn@doi [\aap]
  {10.1051/0004-6361/201322068}, \href
  {http://adsabs.harvard.edu/abs/2013A%26A...558A..33A} {558, A33}

\bibitem[\protect\citeauthoryear{{Bautz} \& {Morgan}}{{Bautz} \&
  {Morgan}}{1970}]{Bautz1970}
{Bautz} L.~P.,  {Morgan} W.~W.,  1970, \mn@doi [\apjl] {10.1086/180643}, \href
  {https://ui.adsabs.harvard.edu/abs/1970ApJ...162L.149B} {162, L149}

\bibitem[\protect\citeauthoryear{Bengio, Simard  \& Frasconi}{Bengio
  et~al.}{1994}]{Bengio1994}
Bengio Y.,  Simard P.,   Frasconi P.,  1994, \mn@doi [Trans. Neur. Netw.]
  {10.1109/72.279181}, 5, 157

\bibitem[\protect\citeauthoryear{{Bertin} \& {Arnouts}}{{Bertin} \&
  {Arnouts}}{1996}]{Bertin96}
{Bertin} E.,  {Arnouts} S.,  1996, \mn@doi [\aaps] {10.1051/aas:1996164}, \href
  {https://ui.adsabs.harvard.edu/abs/1996A%26AS..117..393B} {117, 393}

\bibitem[\protect\citeauthoryear{{Bolya}, {Zhou}, {Xiao}  \& {Lee}}{{Bolya}
  et~al.}{2019}]{Bolya2019}
{Bolya} D.,  {Zhou} C.,  {Xiao} F.,   {Lee} Y.~J.,  2019, arXiv e-prints, \href
  {https://ui.adsabs.harvard.edu/abs/2019arXiv190402689B} {p. arXiv:1904.02689}

\bibitem[\protect\citeauthoryear{{Boucaud} et~al.,}{{Boucaud}
  et~al.}{2019}]{Boucaud19}
{Boucaud} A.,  et~al., 2019, arXiv e-prints, \href
  {https://ui.adsabs.harvard.edu/abs/2019arXiv190501324B} {p. arXiv:1905.01324}

\bibitem[\protect\citeauthoryear{{Burke}, {Aleo}  \& {Chen}}{{Burke}
  et~al.}{2019}]{repo}
{Burke} C.~J.,  {Aleo} P.~D.,   {Chen} Y.-C.,  2019, Astro R-CNN: Instance
  Segmentation in Astronomical Images using Mask R-CNN Deep Learning,
  \url{https://github.com/burke86/astro_rcnn}

\bibitem[\protect\citeauthoryear{{Castelli} \& {Kurucz}}{{Castelli} \&
  {Kurucz}}{2003}]{Castelli2003}
{Castelli} F.,  {Kurucz} R.~L.,  2003, in {Piskunov} N.,  {Weiss} W.~W.,
  {Gray} D.~F.,  eds,  IAU Symposium Vol. 210, Modelling of Stellar
  Atmospheres. p.~A20 (\mn@eprint {arXiv} {astro-ph/0405087})

\bibitem[\protect\citeauthoryear{{Chang} et~al.,}{{Chang}
  et~al.}{2013}]{Chang13}
{Chang} C.,  et~al., 2013, \mn@doi [\mnras] {10.1093/mnras/stt1156}, \href
  {http://adsabs.harvard.edu/abs/2013MNRAS.434.2121C} {434, 2121}

\bibitem[\protect\citeauthoryear{Cheng}{Cheng}{2017}]{Cheng2017}
Cheng J.,  2017, PhD thesis, Purdue University

\bibitem[\protect\citeauthoryear{{Cheng} et~al.,}{{Cheng}
  et~al.}{2019}]{Cheng2019}
{Cheng} T.-Y.,  et~al., 2019, arXiv e-prints, \href
  {https://ui.adsabs.harvard.edu/abs/2019arXiv190803610C} {p. arXiv:1908.03610}

\bibitem[\protect\citeauthoryear{Chollet et~al.}{Chollet
  et~al.}{2015}]{Chollet2015}
Chollet F.,  et~al., 2015, Keras, \url{https://keras.io}

\bibitem[\protect\citeauthoryear{{Couprie} \& {Bertrand}}{{Couprie} \&
  {Bertrand}}{1997}]{Couprie1997}
{Couprie} M.,  {Bertrand} G.,  1997, in {Melter} R.~A.,  {Wu} A.~Y.,
  {Latecki} L.~J.,  eds,  \procspie Vol. 3168, Vision Geometry VI. pp 136--146,
  \mn@doi{10.1117/12.292778}

\bibitem[\protect\citeauthoryear{{D'Isanto} \& {Polsterer}}{{D'Isanto} \&
  {Polsterer}}{2018}]{DIsanto2018}
{D'Isanto} A.,  {Polsterer} K.~L.,  2018, \mn@doi [\aap]
  {10.1051/0004-6361/201731326}, \href
  {https://ui.adsabs.harvard.edu/abs/2018A&A...609A.111D} {609, A111}

\bibitem[\protect\citeauthoryear{Dawson \& Schneider}{Dawson \&
  Schneider}{2014}]{Dawson14}
Dawson W.,  Schneider M.,  2014, Complementarity of LSST and WFIRST: Regarding
  Object Blending

\bibitem[\protect\citeauthoryear{{Dawson}, {Schneider}, {Tyson}  \&
  {Jee}}{{Dawson} et~al.}{2016}]{Dawson16}
{Dawson} W.~A.,  {Schneider} M.~D.,  {Tyson} J.~A.,   {Jee} M.~J.,  2016,
  \mn@doi [\apj] {10.3847/0004-637X/816/1/11}, \href
  {http://adsabs.harvard.edu/abs/2016ApJ...816...11D} {816, 11}

\bibitem[\protect\citeauthoryear{{Dey} et~al.,}{{Dey} et~al.}{2019}]{Dey2019}
{Dey} A.,  et~al., 2019, \mn@doi [\aj] {10.3847/1538-3881/ab089d}, \href
  {https://ui.adsabs.harvard.edu/abs/2019AJ....157..168D} {157, 168}

\bibitem[\protect\citeauthoryear{Dieleman, Willett  \& Dambre}{Dieleman
  et~al.}{2015}]{Dieleman2015}
Dieleman S.,  Willett K.,   Dambre J.,  2015, \mn@doi [\mnras]
  {10.1093/mnras/stv632}, 450, 1441

\bibitem[\protect\citeauthoryear{Everingham, Van~Gool, Williams, Winn  \&
  Zisserman}{Everingham et~al.}{2010}]{Everingham2010}
Everingham M.,  Van~Gool L.,  Williams C. K.~I.,  Winn J.,   Zisserman A.,
  2010, \mn@doi [International Journal of Computer Vision]
  {10.1007/s11263-009-0275-4}, 88, 303

\bibitem[\protect\citeauthoryear{{Flaugher} et~al.,}{{Flaugher}
  et~al.}{2015}]{Flaugher2015}
{Flaugher} B.,  et~al., 2015, \mn@doi [\aj] {10.1088/0004-6256/150/5/150},
  \href {https://ui.adsabs.harvard.edu/abs/2015AJ....150..150F} {150, 150}

\bibitem[\protect\citeauthoryear{{Gallazzi}, {Charlot}, {Brinchmann}, {White}
  \& {Tremonti}}{{Gallazzi} et~al.}{2005}]{Gallazzi2005}
{Gallazzi} A.,  {Charlot} S.,  {Brinchmann} J.,  {White} S. D.~M.,   {Tremonti}
  C.~A.,  2005, \mn@doi [\mnras] {10.1111/j.1365-2966.2005.09321.x}, \href
  {https://ui.adsabs.harvard.edu/abs/2005MNRAS.362...41G} {362, 41}

\bibitem[\protect\citeauthoryear{{Girshick}}{{Girshick}}{2015}]{Girshick2015}
{Girshick} R.,  2015, in 2015 IEEE International Conference on Computer Vision
  (ICCV). pp 1440--1448, \mn@doi{10.1109/ICCV.2015.169}

\bibitem[\protect\citeauthoryear{{Gonz{\'a}lez}, {Mu{\~n}oz}  \&
  {Hern{\'a}ndez}}{{Gonz{\'a}lez} et~al.}{2018}]{Gonzalez2018}
{Gonz{\'a}lez} R.~E.,  {Mu{\~n}oz} R.~P.,   {Hern{\'a}ndez} C.~A.,  2018,
  \mn@doi [Astronomy and Computing] {10.1016/j.ascom.2018.09.004}, \href
  {https://ui.adsabs.harvard.edu/abs/2018A&C....25..103G} {25, 103}

\bibitem[\protect\citeauthoryear{{Hausen} \& {Robertson}}{{Hausen} \&
  {Robertson}}{2019}]{Hausen2019}
{Hausen} R.,  {Robertson} B.,  2019, arXiv e-prints, \href
  {https://ui.adsabs.harvard.edu/abs/2019arXiv190611248H} {p. arXiv:1906.11248}

\bibitem[\protect\citeauthoryear{He, Zhang, Ren  \& Sun}{He
  et~al.}{2016}]{He2016}
He K.,  Zhang X.,  Ren S.,   Sun J.,  2016, 2016 IEEE Conference on Computer
  Vision and Pattern Recognition (CVPR), pp 770--778

\bibitem[\protect\citeauthoryear{He, Gkioxari, Doll\'{a}r  \& Girshick}{He
  et~al.}{2017}]{He17}
He K.,  Gkioxari G.,  Doll\'{a}r P.,   Girshick R.~B.,  2017, 2017 IEEE
  International Conference on Computer Vision (ICCV), pp 2980--2988

\bibitem[\protect\citeauthoryear{Huang, Sun, Liu, Sedra  \& Weinberger}{Huang
  et~al.}{2016}]{Huang2016}
Huang G.,  Sun Y.,  Liu Z.,  Sedra D.,   Weinberger K.,  2016. pp 646--661,
  \mn@doi{10.1007/978-3-319-46493-0_39}

\bibitem[\protect\citeauthoryear{{Huang} et~al.,}{{Huang}
  et~al.}{2018}]{Huang2018}
{Huang} S.,  et~al., 2018, \mn@doi [\pasj] {10.1093/pasj/psx126}, \href
  {https://ui.adsabs.harvard.edu/abs/2018PASJ...70S...6H} {70, S6}

\bibitem[\protect\citeauthoryear{{Huang}, {Huang}, {Gong}, {Huang}  \&
  {Wang}}{{Huang} et~al.}{2019}]{Huang2019}
{Huang} Z.,  {Huang} L.,  {Gong} Y.,  {Huang} C.,   {Wang} X.,  2019, arXiv
  e-prints, \href {https://ui.adsabs.harvard.edu/abs/2019arXiv190300241H} {p.
  arXiv:1903.00241}

\bibitem[\protect\citeauthoryear{Hunter}{Hunter}{2007}]{Hunter2007}
Hunter J.~D.,  2007, \mn@doi [Computing in Science \& Engineering]
  {10.1109/MCSE.2007.55}, 9, 90

\bibitem[\protect\citeauthoryear{{Iglovikov} \& {Shvets}}{{Iglovikov} \&
  {Shvets}}{2018}]{Iglovikov2018}
{Iglovikov} V.,  {Shvets} A.,  2018, arXiv e-prints, \href
  {https://ui.adsabs.harvard.edu/abs/2018arXiv180105746I} {p. arXiv:1801.05746}

\bibitem[\protect\citeauthoryear{{Ivezi{\'c}} et~al.,}{{Ivezi{\'c}}
  et~al.}{2019}]{Ivezic2019}
{Ivezi{\'c}} {\v{Z}}.,  et~al., 2019, \mn@doi [\apj]
  {10.3847/1538-4357/ab042c}, \href
  {https://ui.adsabs.harvard.edu/abs/2019ApJ...873..111I} {873, 111}

\bibitem[\protect\citeauthoryear{{Jarvis} \& {Tyson}}{{Jarvis} \&
  {Tyson}}{1981}]{Jarvis81}
{Jarvis} J.~F.,  {Tyson} J.~A.,  1981, \mn@doi [\aj] {10.1086/112907}, \href
  {https://ui.adsabs.harvard.edu/abs/1981AJ.....86..476J} {86, 476}

\bibitem[\protect\citeauthoryear{{Kim} \& {Brunner}}{{Kim} \&
  {Brunner}}{2017}]{Kim2017}
{Kim} E.~J.,  {Brunner} R.~J.,  2017, \mn@doi [\mnras] {10.1093/mnras/stw2672},
  \href {https://ui.adsabs.harvard.edu/abs/2017MNRAS.464.4463K} {464, 4463}

\bibitem[\protect\citeauthoryear{{Kim}, {Brunner}  \& {Carrasco Kind}}{{Kim}
  et~al.}{2015}]{Kim2015}
{Kim} E.~J.,  {Brunner} R.~J.,   {Carrasco Kind} M.,  2015, \mn@doi [\mnras]
  {10.1093/mnras/stv1608}, \href
  {https://ui.adsabs.harvard.edu/abs/2015MNRAS.453..507K} {453, 507}

\bibitem[\protect\citeauthoryear{{Kingma} \& {Ba}}{{Kingma} \&
  {Ba}}{2014}]{Kingma2014}
{Kingma} D.~P.,  {Ba} J.,  2014, arXiv e-prints, \href
  {https://ui.adsabs.harvard.edu/abs/2014arXiv1412.6980K} {p. arXiv:1412.6980}

\bibitem[\protect\citeauthoryear{Krizhevsky, Sutskever  \& Hinton}{Krizhevsky
  et~al.}{2012}]{Krizhevsky2012}
Krizhevsky A.,  Sutskever I.,   Hinton G.~E.,  2012, in Pereira F.,  Burges C.
  J.~C.,  Bottou L.,   Weinberger K.~Q.,  eds, , Advances in Neural Information
  Processing Systems 25.
Curran Associates, Inc., pp 1097--1105

\bibitem[\protect\citeauthoryear{{Kroupa}}{{Kroupa}}{2001}]{Kroupa2001}
{Kroupa} P.,  2001, \mn@doi [\mnras] {10.1046/j.1365-8711.2001.04022.x}, \href
  {https://ui.adsabs.harvard.edu/abs/2001MNRAS.322..231K} {322, 231}

\bibitem[\protect\citeauthoryear{{Kurucz}}{{Kurucz}}{1993}]{Kurucz1993}
{Kurucz} R.,  1993, ATLAS9 Stellar Atmosphere Programs and 2 km/s grid.~Kurucz
  CD-ROM No.~13.~ Cambridge, Mass.: Smithsonian Astrophysical Observatory,
  1993., \href {https://ui.adsabs.harvard.edu/abs/1993KurCD..13.....K} {13}

\bibitem[\protect\citeauthoryear{Lin, Maire, Belongie, Hays, Perona, Ramanan,
  Doll{\'a}r  \& Zitnick}{Lin et~al.}{2014}]{Lin2014}
Lin T.-Y.,  Maire M.,  Belongie S.,  Hays J.,  Perona P.,  Ramanan D.,
  Doll{\'a}r P.,   Zitnick C.~L.,  2014, in European Conference on Computer
  Vision (ECCV). Z{\"u}rich

\bibitem[\protect\citeauthoryear{Lin, Doll{\'a}r, Girshick, He, Hariharan  \&
  Belongie}{Lin et~al.}{2017}]{Lin17}
Lin T.-Y.,  Doll{\'a}r P.,  Girshick R.~B.,  He K.,  Hariharan B.,   Belongie
  S.~J.,  2017, 2017 IEEE Conference on Computer Vision and Pattern Recognition
  (CVPR), pp 936--944

\bibitem[\protect\citeauthoryear{{Liu}, {Qi}, {Qin}, {Shi}  \& {Jia}}{{Liu}
  et~al.}{2018}]{Liu2018}
{Liu} S.,  {Qi} L.,  {Qin} H.,  {Shi} J.,   {Jia} J.,  2018, arXiv e-prints,
  \href {https://ui.adsabs.harvard.edu/abs/2018arXiv180301534L} {p.
  arXiv:1803.01534}

\bibitem[\protect\citeauthoryear{{Lupton}}{{Lupton}}{2014}]{Lupton2014}
{Lupton} R.,  2014, {Joint Source Detection, Deblending, and Measurement for
  WFIRST-AFTA and LSST}, NASA WPS Proposal

\bibitem[\protect\citeauthoryear{{Lupton}, {Blanton}, {Fekete}, {Hogg},
  {O'Mullane}, {Szalay}  \& {Wherry}}{{Lupton} et~al.}{2004}]{Lupton2004}
{Lupton} R.,  {Blanton} M.~R.,  {Fekete} G.,  {Hogg} D.~W.,  {O'Mullane} W.,
  {Szalay} A.,   {Wherry} N.,  2004, \mn@doi [\pasp] {10.1086/382245}, \href
  {https://ui.adsabs.harvard.edu/abs/2004PASP..116..133L} {116, 133}

\bibitem[\protect\citeauthoryear{{Madau} \& {Dickinson}}{{Madau} \&
  {Dickinson}}{2014}]{Madau2014}
{Madau} P.,  {Dickinson} M.,  2014, \mn@doi [\araa]
  {10.1146/annurev-astro-081811-125615}, \href
  {https://ui.adsabs.harvard.edu/abs/2014ARA&A..52..415M} {52, 415}

\bibitem[\protect\citeauthoryear{{Melchior}, {Moolekamp}, {Jerdee},
  {Armstrong}, {Sun}, {Bosch}  \& {Lupton}}{{Melchior}
  et~al.}{2018}]{Melchior2018}
{Melchior} P.,  {Moolekamp} F.,  {Jerdee} M.,  {Armstrong} R.,  {Sun} A.~L.,
  {Bosch} J.,   {Lupton} R.,  2018, \mn@doi [Astronomy and Computing]
  {10.1016/j.ascom.2018.07.001}, \href
  {https://ui.adsabs.harvard.edu/abs/2018A&C....24..129M} {24, 129}

\bibitem[\protect\citeauthoryear{{Messier}}{{Messier}}{1781}]{Messier1781}
{Messier} C.,  1781, Technical report, {Catalogue des N{\'e}buleuses et des
  Amas d'{\'E}toiles (Catalog of Nebulae and Star Clusters)}

\bibitem[\protect\citeauthoryear{{Moll{\'a}}, {Garc{\'{\i}}a-Vargas}  \&
  {Bressan}}{{Moll{\'a}} et~al.}{2009}]{Molla2009}
{Moll{\'a}} M.,  {Garc{\'{\i}}a-Vargas} M.~L.,   {Bressan} A.,  2009, \mn@doi
  [\mnras] {10.1111/j.1365-2966.2009.15160.x}, \href
  {https://ui.adsabs.harvard.edu/abs/2009MNRAS.398..451M} {398, 451}

\bibitem[\protect\citeauthoryear{{Olowin}}{{Olowin}}{1988}]{Olowin1988}
{Olowin} R.~P.,  1988, \mn@doi [\pasp] {10.1086/132333}, \href
  {https://ui.adsabs.harvard.edu/abs/1988PASP..100.1354O} {100, 1354}

\bibitem[\protect\citeauthoryear{{Pasquet}, {Bertin}, {Treyer}, {Arnouts}  \&
  {Fouchez}}{{Pasquet} et~al.}{2019}]{Pasquet2019}
{Pasquet} J.,  {Bertin} E.,  {Treyer} M.,  {Arnouts} S.,   {Fouchez} D.,  2019,
  \mn@doi [\aap] {10.1051/0004-6361/201833617}, \href
  {https://ui.adsabs.harvard.edu/abs/2019A&A...621A..26P} {621, A26}

\bibitem[\protect\citeauthoryear{{Peletier} \& {Balcells}}{{Peletier} \&
  {Balcells}}{1996}]{Peletier1996}
{Peletier} R.~F.,  {Balcells} M.,  1996, \mn@doi [\aj] {10.1086/117958}, \href
  {https://ui.adsabs.harvard.edu/abs/1996AJ....111.2238P} {111, 2238}

\bibitem[\protect\citeauthoryear{{Pence}, {Chiappetti}, {Page}, {Shaw}  \&
  {Stobie}}{{Pence} et~al.}{2010}]{Pence2010}
{Pence} W.~D.,  {Chiappetti} L.,  {Page} C.~G.,  {Shaw} R.~A.,   {Stobie} E.,
  2010, \mn@doi [\aap] {10.1051/0004-6361/201015362}, \href
  {https://ui.adsabs.harvard.edu/abs/2010A&A...524A..42P} {524, A42}

\bibitem[\protect\citeauthoryear{{Peterson} et~al.,}{{Peterson}
  et~al.}{2015}]{Peterson2015}
{Peterson} J.~R.,  et~al., 2015, \mn@doi [\apjs] {10.1088/0067-0049/218/1/14},
  \href {https://ui.adsabs.harvard.edu/abs/2015ApJS..218...14P} {218, 14}

\bibitem[\protect\citeauthoryear{{Price-Whelan} et~al.,}{{Price-Whelan}
  et~al.}{2018}]{astropy:2018}
{Price-Whelan} A.~M.,  et~al., 2018, \mn@doi [\aj] {10.3847/1538-3881/aabc4f},
  \href {https://ui.adsabs.harvard.edu/#abs/2018AJ....156..123T} {156, 123}

\bibitem[\protect\citeauthoryear{{Prugniel}, {Vauglin}  \& {Koleva}}{{Prugniel}
  et~al.}{2011}]{Prugniel2011}
{Prugniel} P.,  {Vauglin} I.,   {Koleva} M.,  2011, \mn@doi [\aap]
  {10.1051/0004-6361/201116769}, \href
  {https://ui.adsabs.harvard.edu/abs/2011A&A...531A.165P} {531, A165}

\bibitem[\protect\citeauthoryear{{Redmon}, {Divvala}, {Girshick}  \&
  {Farhadi}}{{Redmon} et~al.}{2015}]{Redmon2015}
{Redmon} J.,  {Divvala} S.,  {Girshick} R.,   {Farhadi} A.,  2015, arXiv
  e-prints, \href {https://ui.adsabs.harvard.edu/abs/2015arXiv150602640R} {p.
  arXiv:1506.02640}

\bibitem[\protect\citeauthoryear{Reiman \& G\"{o}hre}{Reiman \&
  G\"{o}hre}{2019}]{Reiman19}
Reiman D.~M.,  G\"{o}hre B.~E.,  2019, \mn@doi [\mnras] {10.1093/mnras/stz575},
  485, 2617

\bibitem[\protect\citeauthoryear{Ren, He, Girshick  \& Sun}{Ren
  et~al.}{2015}]{Ren2015}
Ren S.,  He K.,  Girshick R.,   Sun J.,  2015, in Cortes C.,  Lawrence N.~D.,
  Lee D.~D.,  Sugiyama M.,   Garnett R.,  eds, , Advances in Neural Information
  Processing Systems 28.
Curran Associates, Inc., pp 91--99

\bibitem[\protect\citeauthoryear{Ronneberger, Fischer  \& Brox}{Ronneberger
  et~al.}{2015}]{Ronneberger2015}
Ronneberger O.,  Fischer P.,   Brox T.,  2015, in MICCAI.

\bibitem[\protect\citeauthoryear{{Sebok}}{{Sebok}}{1979}]{Sebok1979}
{Sebok} W.~L.,  1979, \mn@doi [\aj] {10.1086/112570}, \href
  {https://ui.adsabs.harvard.edu/abs/1979AJ.....84.1526S} {84, 1526}

\bibitem[\protect\citeauthoryear{{Serra-Ricart}, {Gaitan}, {Garrido}  \&
  {Perez-Fournon}}{{Serra-Ricart} et~al.}{1996}]{Serra-Ricart1996}
{Serra-Ricart} M.,  {Gaitan} V.,  {Garrido} L.,   {Perez-Fournon} I.,  1996,
  \aaps, \href {https://ui.adsabs.harvard.edu/abs/1996A%26AS..115..195S} {115,
  195}

\bibitem[\protect\citeauthoryear{{S{\'e}rsic}}{{S{\'e}rsic}}{1963}]{Sersic1963}
{S{\'e}rsic} J.~L.,  1963, Boletin de la Asociacion Argentina de Astronomia La
  Plata Argentina, \href
  {https://ui.adsabs.harvard.edu/abs/1963BAAA....6...41S} {6, 41}

\bibitem[\protect\citeauthoryear{{Sevilla-Noarbe} et~al.,}{{Sevilla-Noarbe}
  et~al.}{2018}]{Sevilla-Noarbe2018}
{Sevilla-Noarbe} I.,  et~al., 2018, \mn@doi [\mnras] {10.1093/mnras/sty2579},
  \href {https://ui.adsabs.harvard.edu/abs/2018MNRAS.481.5451S} {481, 5451}

\bibitem[\protect\citeauthoryear{{Soumagnac} et~al.,}{{Soumagnac}
  et~al.}{2015}]{Soumagnac2015}
{Soumagnac} M.~T.,  et~al., 2015, \mn@doi [\mnras] {10.1093/mnras/stu1410},
  \href {https://ui.adsabs.harvard.edu/abs/2015MNRAS.450..666S} {450, 666}

\bibitem[\protect\citeauthoryear{{Spergel} et~al.,}{{Spergel}
  et~al.}{2013}]{Spergel13}
{Spergel} D.,  et~al., 2013, arXiv e-prints, \href
  {https://ui.adsabs.harvard.edu/abs/2013arXiv1305.5422S} {p. arXiv:1305.5422}

\bibitem[\protect\citeauthoryear{Tan, Sun, Kong, Zhang, Yang  \& Liu}{Tan
  et~al.}{2018}]{Tan2018}
Tan C.,  Sun F.,  Kong T.,  Zhang W.,  Yang C.,   Liu C.,  2018, A Survey on
  Deep Transfer Learning: 27th International Conference on Artificial Neural
  Networks, Rhodes, Greece, October 4-7, 2018, Proceedings, Part III.
pp 270--279, \mn@doi{10.1007/978-3-030-01424-7_27}

\bibitem[\protect\citeauthoryear{{Taylor}, {Dye}, {Broadhurst},
  {Ben{\'{\i}}tez}  \& {van Kampen}}{{Taylor} et~al.}{1998}]{Taylor1998}
{Taylor} A.~N.,  {Dye} S.,  {Broadhurst} T.~J.,  {Ben{\'{\i}}tez} N.,   {van
  Kampen} E.,  1998, \mn@doi [\apj] {10.1086/305827}, \href
  {https://ui.adsabs.harvard.edu/abs/1998ApJ...501..539T} {501, 539}

\bibitem[\protect\citeauthoryear{Tsai, Gajda, Sloan, Rares  \& Shen}{Tsai
  et~al.}{2019}]{Tsai2019}
Tsai H.-F.,  Gajda J.,  Sloan T.~F.,  Rares A.,   Shen A.~Q.,  2019, \mn@doi
  [SoftwareX] {https://doi.org/10.1016/j.softx.2019.02.007}, 9, 230

\bibitem[\protect\citeauthoryear{{Tyson} \& {Fischer}}{{Tyson} \&
  {Fischer}}{1995}]{Tyson1995}
{Tyson} J.~A.,  {Fischer} P.,  1995, \mn@doi [\apjl] {10.1086/187929}, \href
  {https://ui.adsabs.harvard.edu/abs/1995ApJ...446L..55T} {446, L55}

\bibitem[\protect\citeauthoryear{{Valdes}}{{Valdes}}{1982}]{Valdes1982}
{Valdes} F.,  1982, in Instrumentation in Astronomy IV. pp 465--472,
  \mn@doi{10.1117/12.933489}

\bibitem[\protect\citeauthoryear{{Vasconcellos}, {de Carvalho}, {Gal},
  {LaBarbera}, {Capelato}, {Frago Campos Velho}, {Trevisan}  \&
  {Ruiz}}{{Vasconcellos} et~al.}{2011}]{Vasconellos2011}
{Vasconcellos} E.~C.,  {de Carvalho} R.~R.,  {Gal} R.~R.,  {LaBarbera} F.~L.,
  {Capelato} H.~V.,  {Frago Campos Velho} H.,  {Trevisan} M.,   {Ruiz}
  R.~S.~R.,  2011, \mn@doi [\aj] {10.1088/0004-6256/141/6/189}, \href
  {https://ui.adsabs.harvard.edu/abs/2011AJ....141..189V} {141, 189}

\bibitem[\protect\citeauthoryear{{Zhang} \& {Bloom}}{{Zhang} \&
  {Bloom}}{2019}]{Zhang2019}
{Zhang} K.,  {Bloom} J.~S.,  2019, arXiv e-prints, \href
  {https://ui.adsabs.harvard.edu/abs/2019arXiv190709500Z} {p. arXiv:1907.09500}

\bibitem[\protect\citeauthoryear{{Zhang}, {McKay}, {Bertin}, {Jeltema},
  {Miller}, {Rykoff}  \& {Song}}{{Zhang} et~al.}{2015}]{Zhang2015}
{Zhang} Y.,  {McKay} T.~A.,  {Bertin} E.,  {Jeltema} T.,  {Miller} C.~J.,
  {Rykoff} E.,   {Song} J.,  2015, \mn@doi [\pasp] {10.1086/684053}, \href
  {https://ui.adsabs.harvard.edu/abs/2015PASP..127.1183Z} {127, 1183}

\bibitem[\protect\citeauthoryear{Zhang, Witharana, Liljedahl  \&
  Kanevskiy}{Zhang et~al.}{2018}]{Zhang2018}
Zhang W.,  Witharana C.,  Liljedahl A.~K.,   Kanevskiy M.,  2018, \mn@doi
  [Remote Sensing] {10.3390/rs10091487}, 10

\bibitem[\protect\citeauthoryear{{Zimmermann} \& {Siems}}{{Zimmermann} \&
  {Siems}}{2018}]{Zimmermann2018}
{Zimmermann} R.~S.,  {Siems} J.~N.,  2018, arXiv e-prints, \href
  {https://ui.adsabs.harvard.edu/abs/2018arXiv180907069Z} {p. arXiv:1809.07069}

\makeatother
\end{thebibliography}




\appendix

\section{Loss Curves}
\label{sec:loss}

\begin{figure}
\includegraphics[width=\columnwidth]{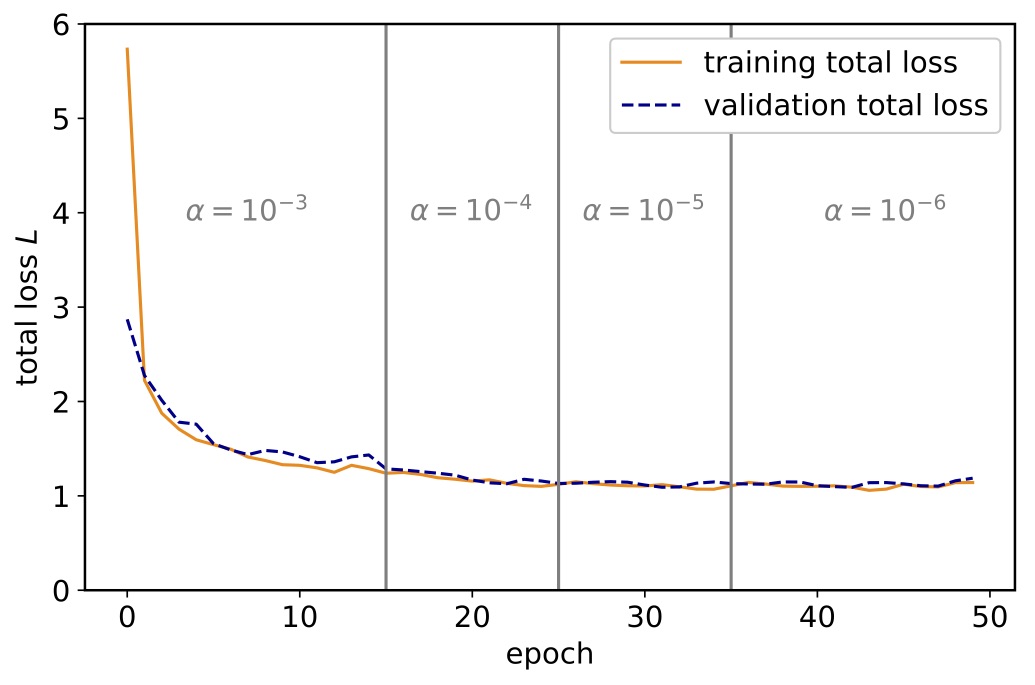}
    \caption{Mask R-CNN total loss versus training epoch.}
    \label{fig:total_loss}
\includegraphics[width=\columnwidth]{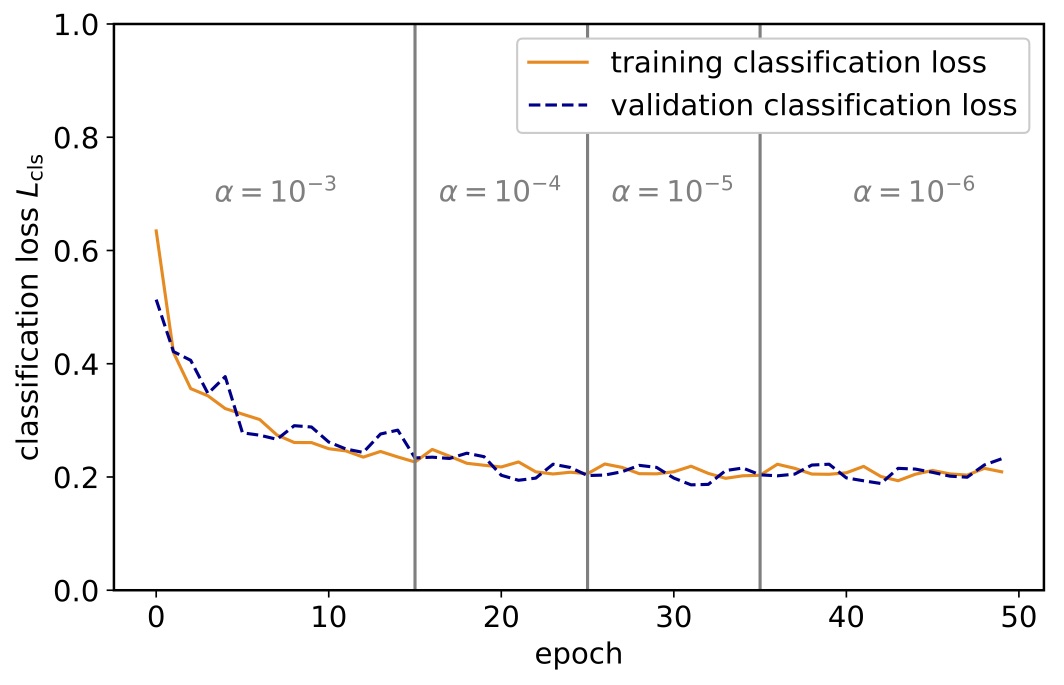}
    \caption{Mask R-CNN class loss versus training epoch.}
    \label{fig:class_loss}
\includegraphics[width=\columnwidth]{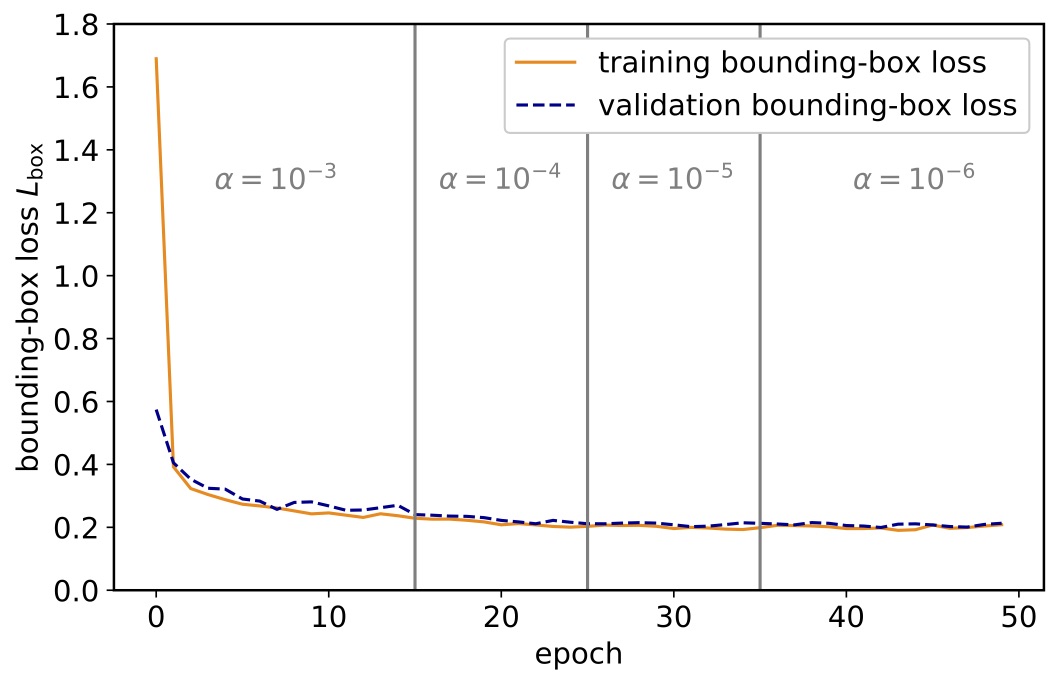}
    \caption{Mask R-CNN bounding box loss versus training epoch.}
    \label{fig:bbox_loss}
\end{figure}

\begin{figure}
\includegraphics[width=\columnwidth]{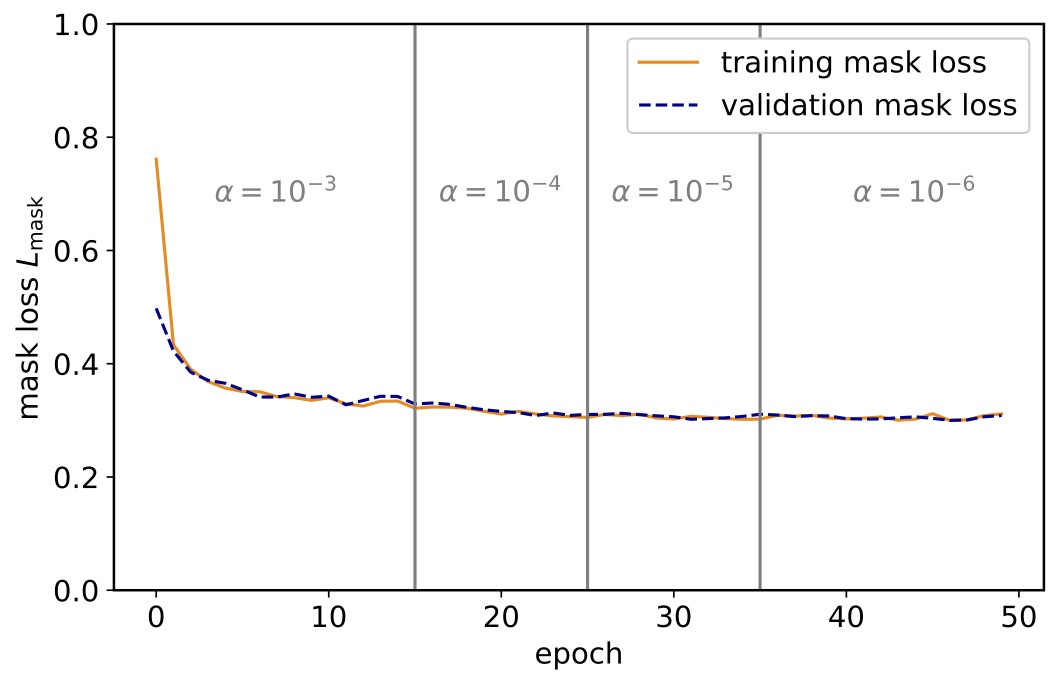}
    \caption{Mask R-CNN mask loss versus training epoch.}
    \label{fig:mask_loss}
\end{figure}

We show the training loss versus epoch following the learning schedule described in \S\ref{sec:training} in Figures~\ref{fig:total_loss},\ref{fig:class_loss},\ref{fig:bbox_loss},\ref{fig:mask_loss}. The total loss $L$ is defined as the sum of the class, bounding box, and mask losses: $L = L_\text{cls} + L_\text{box} + L_\text{mask}$ \citep{He17}. We stop training at 50 epochs. After $\sim$50 epochs, the asymptotic loss curves show diminishing returns. We note that the validation loss is consistent with the training loss. Scheduling the learning rate to progressively decrease during the training enables refinement of the network performance without overfitting on the training set.


\bsp	
\label{lastpage}
\end{document}